\documentclass{aa}  

\usepackage{graphicx}
\usepackage{txfonts}
\usepackage{pdflscape}
\usepackage{afterpage}
\usepackage{natbib}
\usepackage[normalem]{ulem}
\usepackage{rotating}
\usepackage{float}

\makeatletter
\let\linenumbers\relax
\let\endlinenumbers\relax
\let\switchlinenumbers\relax
\let\modulolinenumbers\relax
\makeatother

\usepackage{censor}
\newlength\nextcharwidth
\makeatletter
\renewcommand\@cenword[1]{%
  \setlength{\nextcharwidth}{\widthof{#1}}%
  \censorrule{\nextcharwidth}%
  \kern -\nextcharwidth%
  #1}
\makeatother

\usepackage[colorlinks=true,linkcolor=blue,citecolor=blue]{hyperref}
\begin{document} 
\defcitealias{NDiaye2024}{Paper~I}
\title{Cascade adaptive optics with a second stage based on a Zernike wavefront sensor for exoplanet observations\\
II. Validation in broadband light on the ESO/GHOST testbed}

\titlerunning{Cascade AO with ZWFS – Part II: Validation in broadband light}
\authorrunning{A. Rahim et al.}

\author{A. Rahim\inst{1} \and
M. N’Diaye\inst{1} \and
A. Vigan\inst{2} \and
M. Kasper\inst{3} \and
J. Nousiainen\inst{3}\and
B. Engler\inst{3} \and
K. Dohlen\inst{2} \and
S. Leveratto\inst{3} \and\\
J. Floriot\inst{2} \and
M. Marcos\inst{2} \and
C. Bailet\inst{1} \and
P. Bristow\inst{3} \and
E. S. Douglas\inst{4}
 }

\institute{Université Côte d’Azur, Observatoire de la Côte d’Azur, CNRS, Laboratoire Lagrange, France\\
             \email{anna.rahim@oca.eu}
            \and Aix Marseille Université, CNRS, CNES, LAM, Marseille, France
            \and European Southern Observatory (ESO), Karl-Schwarzschild-Str. 2, 85748 Garching, Germany
            \and Department of Astronomy and Steward Observatory, University of Arizona, Tucson, AZ 85721, USA}
   \date{}

 
  \abstract
   {Current high-contrast facilities on the ground use extreme adaptive optics (XAO) systems
   to achieve contrasts down to $10^{-6}$ at 200\,mas for exoplanet observations. This performance is mainly limited by the XAO residuals due to the temporal errors in the XAO control loop. To overcome this issue, a promising solution consists in using cascade adaptive optics with a fast second stage. This approach was recently validated for a control loop based on a Zernike wavefront sensor (ZWFS) in monochromatic light.}
   {As wavefront sensors operate in broadband light to maximise photon sensitivity, this work aims to validate the ZWFS-based control loop in polychromatic light and assess its performance over a wide range of seeings, wind speeds, and stellar fluxes.}
   {Experiments were conducted on the ESO's GPU-based High-order adaptive OpticS Testbench (GHOST) testbed to probe our scheme in polychromatic light. Residual aberrations from a first-stage XAO system were simulated and our approach was evaluated in narrowband and broadband light through contrast in Lyot coronagraphic images. }
   {Closing the ZWFS-based control loop consistently improves contrast within its correction region in most tested conditions. After subtraction of quasi-static aberrations, our loop reaches a contrast gain up to one order of magnitude, independently of bandwidth and turbulence strength. The broadband and narrowband cases match in performance for bright sources, while narrowband remains slightly preferable for faint targets. }
   {These results demonstrate the feasibility of broadband ZWFS-based control loop and underline promising avenues with achromatic masks and an accurate calibration of quasi-static aberrations for future high-contrast imaging on Extremely Large Telescopes. 
}


   \maketitle
%

\section{Introduction}
Understanding the formation and evolution of planetary systems is nowadays one of the key challenges in astrophysics. Direct imaging is a promising approach to provide clues and address this topic, as it both enables the detection of exoplanets and the characterisation of their atmospheres. This technique is inherently limited by the combination of low planet-to-star flux ratios at small angular separations, with contrasts typically ranging from $10^{-4}$ to $10^{-10}$ at distances ranging from a few hundreds to a few tens of milliarcseconds in the visible and near infrared \citep[e.g.,][]{Currie2023,Chauvin2024}. These constraints impose severe requirements on high-contrast instrumentation for two complementary classes of facilities: space missions with coronagraphic capabilities and ground-based extremely large telescopes (ELTs).

Space-based observatories such as the Nancy Grace Roman Space Telescope (RST) and the envisioned Habitable Worlds Observatory (HWO) will operate in an exceptionally stable environment free from atmospheric turbulence. These missions target planet-to-star contrasts smaller than 10$^{-8}$ through a combination of ultra-stable telescopes and coronagraphs with high-precision wavefront control \citep{Bailey2023,Mennesson2024}. In parallel, ground-based facilities such as the Extremely Large Telescope (ELT), the Giant Magellan Telescope (GMT), and the Thirty Meter Telescope (TMT) with high-contrast capabilities will benefit from large primary mirrors of 25 to 39\,m diameter, providing unprecedented angular resolution and photon collecting power \citep{Kasper2021,Burgett2024,Fitzgerald2019}. Their apertures typically exceed those of space telescopes (typically 2 to 8\,m) since ground-based facilities are not constrained by launch fairing dimensions, mass limits, or the extreme mechanical and thermal stability required for space operation. However, telescopes on the ground suffer from the impact of the atmospheric turbulence on the image quality. Equipped with extreme adaptive optics (XAO) systems to correct for these effects, these telescopes aim to deliver images with Strehl ratios larger than 90\% in the near infrared \citep[e.g.,][]{Guyon2018}. However, even state-of-the-art XAO instruments on 8\,m-class telescopes such as Gemini/GPI, Subaru/SCExAO, and VLT/SPHERE remain limited by residual temporal wavefront errors, linked to correction speed, restricting achievable contrasts to the 10$^{-4}$-10$^{-6}$ range at 200\,mas \citep{Macintosh2014,Jovanovic2015,Beuzit2019}.

Given these limitations, several strategies have been proposed to further reduce the residual aberrations left by the XAO loop. One promising approach is to cascade adaptive optics (AO) systems, so that a second, more sensitive control stage operates downstream of the main XAO system to correct for the remaining small-amplitude wavefront errors \citep[e.g.,][]{CerpaUrra2022,NDiaye2024}. To measure these residual aberrations, the second stage requires a wavefront sensor with excellent sensitivity. Among the possible options, the Pyramid Wavefront Sensor \citep[PWFS,][]{Ragazzoni1996} and the Zernike Wavefront Sensor \citep[ZWFS,][]{Zernike1934} are two of the leading candidates for this role \citep{Guyon2005}.

The PWFS can be operated in two distinct regimes: modulated or unmodulated. Modulation increases the dynamic range of the sensor at the expense of sensitivity \citep{Ragazzoni1999}. For this reason, many current PWFS implementations rely on modulated operation, which facilitates its use under operational conditions but introduces additional complexity, both from an opto-mechanical and a calibration standpoint. This approach is currently being considered in the context of SAXO+, the second-stage adaptive optics system under development for SPHERE \citep{Boccaletti2022a}. While modulation is effective in extending the dynamic range, it may prove less relevant for a second-stage correction loop in the presence of low amplitude residual aberrations.

The unmodulated PWFS is theoretically very sensitive, but in its standard four-face implementation, it requires the acquisition and processing of four pupil images. This increases the complexity of the data processing chain, as it demands a larger detector area and longer processing times compared to approaches based on a single pupil image. In addition, the unmodulated PWFS presents several limitations, including a light distribution that strongly depends on the incoming phase, a limited intrinsic capture range in the absence of modulation and an optical architecture involving multiple paths and tight alignment tolerances \citep{Chambouleyron2021_pwfs,Agapito2023}.

In contrast, the ZWFS operates with a single pupil image, enabling simple and fast reconstruction. Its response is highly linear in the small aberration regime, and its moderate capture range is well-suited for the measurement of low-amplitude residuals expected after a first stage XAO correction. From an implementation standpoint, the ZWFS relies only on a simple focal-plane phase mask, making a control loop based on the ZWFS compact, stable, and easy to integrate. These characteristics combined with its optimal photon-noise sensitivity \citep[e.g.,][]{Guyon2005,Ndiaye2013,Chambouleyron2021} explain the emergence of the ZWFS  as a particularly attractive solution for driving the second-stage control loop in cascade adaptive optics architectures.

In this context, \citet{NDiaye2024}, hereafter \citetalias{NDiaye2024}, proposed a second-stage adaptive optics control loop based on a ZWFS and carried out its experimental validation on Graphics Processing Unit (GPU)-based High-order adaptive OpticS Testbench (GHOST), the ESO testbed at Garching, Germany \citep{Engler2024_GHOST}. This study successfully demonstrated the operation of the control loop in the small aberration regime, thereby experimentally confirming the potential of the ZWFS-based second stage for fine wavefront error correction in cascade adaptive optics architectures.

However, this demonstration was performed in monochromatic light at the wavelength  $\lambda = 770$\,nm, chosen as a simplified experimental framework for a first proof of concept and validation of the loop operation. Future high-contrast instruments will most likely operate in broadband light with their wavefront-sensing channel to maximise photon throughput and improve wavefront correction performance. Broadband operation introduces additional chromatic challenges, such as wavelength-dependent sensitivity of the Zernike mask and the chromatic behaviour of upstream optics, which may affect loop stability and contrast performance.

The objective of the present work is to assess the behaviour of a ZWFS-driven second stage AO loop under controlled broadband light. Using the GHOST testbed, we emulate realistic XAO residuals and evaluate the loop performance under various spectral bandwidths, turbulence strengths, and atmospheric conditions.

\section{Theory of the Zernike Wavefront Sensor}
High-performance cascade AO requires a wavefront sensor that combines high sensitivity with small non-common path errors to achieve optimal wavefront error correction. Such a performance is essential to reach sufficiently deep contrast levels, which are a prerequisite for the detection of exoplanets with extremely faint signals. In this section, we recall the main properties of the ZWFS and describe the specific limitations that arise when operating in broadband light, following the formalism of \citet{Haffert2024}.

\subsection{Monochromatic light case}
We here assume the observation of an unresolved on-axis star with a telescope of diameter $D$ and no turbulence. The wavefront from the star is assumed to be flat at the telescope entrance pupil. At the telescope focus, we obtain the image of the source which corresponds to the response of the optical system, the so-called Point-Spread Function (PSF).  
The ZWFS is a common path interferometric wavefront sensor in which the central part of the electric field in the stellar PSF is used as a reference field after being phase-shifted by typically $\pi/2$ with a focal-plane mask of relative size $d$ of about one resolution element \citep{Ndiaye2013, Haffert2024}.

The incoming electric field $E_i$ can be expressed as $E_i = A_i e^{i\phi_i}$, where $A_i$ and $\phi_i$ denote the incident amplitude and phase. The focal plane mask introduces a phase shift $\theta$ in the electric field at the core of the PSF thereby generating a reference field $E_r$, defined as $E_r = A_r e^{i\phi_r}$, with $A_r$ and $\phi_r$ denoting the amplitude and phase of the reference field.

The electric field $E_o$ in the output pupil after the mask is given by
\begin{equation}
    E_o = E_i + \mathcal{F}\left\{\mathcal{F}(E_i)\,(M - 1)\right\} = E_i + E_r\,,
\end{equation}
where $\mathcal{F}$ denotes the Fourier transform operator and $M$ is the phase mask.

The resulting intensity $I_o$ in the pupil plane is $I_o=|E_o|^2$ which leads to
\begin{equation}
    I_o = |E_i|^2 + |E_r|^2 + 
    2\,|E_i|\,|E_r|\cos(\phi_i - \phi_r)\,.
\end{equation}

Since detectors only measure the intensity of the electric field, the phase of the wavefront has to be inferred indirectly from the resulting interference pattern. 

This interferometric behaviour of the ZWFS can be further detailed by analysing the dependance of $|E_r|$ and $\phi_r$ on the Zernike mask parameters $d$ and $\theta$, for a clear circular pupil without loss of generality. For a given $\theta$, as $d$ increases, $\phi_r$ remains nearly invariant at each point of the entire pupil. In contrast, the peak of $|E_r|$ increases since a larger dimple captures more light of the PSF in the focal plane. For a given $d$, $\phi_r$ varies with $\theta$ as it is directly linked to the phase delay introduced by the mask. At the same time, the peak of $|E_r|$ also varies with $\theta$ since the light contributions that passes through and surrounds the dimple are both evolving, which leads to a modification of the amplitude of the reference field. To sum up, the mask diameter mainly alters $|E_r|$, while the mask phase delay has an impact on both $|E_r|$ and $\phi_r$.

In the regime of small wavefront aberrations, a linear phase estimator can be obtained by performing a Taylor expansion of the incoming phase around a flat wavefront. This approach provides a straightforward relationship between the measured intensities and the incident phase \citep{Haffert2024} with
\begin{equation}
    \phi_i \approx 
    -\frac{1}{\sin(\phi_r)} \,
    \frac{I_o - I_i - I_r}{2\sqrt{I_i I_r}}\,.
\end{equation}
Furthermore, this linear model can be applied iteratively, after each correction, the wavefront is brought closer to the reference state, maintaining the validity of the linear approximation. The reference electric field is updated, shifting the point around which the linearisation is performed, without the need to explicitly recompute the Taylor expansion.

In its standard version (d=1.06\,$\lambda/D$, $\theta=\pi/2$), the ZWFS provides near optimal photon noise sensitivity 
\citep{Guyon2005, Chambouleyron2021}, but its  mask limits the linear range of the phase to about 0.25\,rad Root Mean Square (RMS). In addition, the ZWFS exhibits an intrinsically asymmetric linear range around the zero point.

Several strategies have been proposed to address these limitations. Phase-shifting approaches, which acquire several images with different phase delays, extend the usable range, thereby improving symmetry in the reconstructed response \citep{Haffert2024}. Vector ZWFS implementations achieve intrinsic symmetry in the capture range by applying two opposite phase shifts simultaneously through polarization‑selective elements \citep{Doelman2019}. In addition, multi-chromatic and liquid-crystal-based techniques further increase the effective capture range, reaching values of at least $ 0.75$rad RMS \citep[e.g.,][]{Wallace2011, Doelman2019, Haffert2024}.
Finally, increasing the mask diameter to $d = 2,\lambda/D$  provides even better photon‑noise sensitivity for all the modes except the low‑order aberrations such as tip‑tilt, as shown by \citet{Chambouleyron2021}.

In the present work, such approaches are not investigated as we consider a second-stage control loop designed to correct for the small amplitude residual aberrations left by a first stage XAO system that mimics the behaviour of a SPHERE-like instrument.

\subsection{Broadband light case}

Using the ZWFS with broadband light introduces two main chromatic effects that reduce both sensitivity and linearity.

First, the physical depth $z$ of the mask induces a wavelength-dependent phase shift $\theta$ with
\begin{equation}
    \theta(\lambda) = 
    \frac{2\pi}{\lambda}(n - 1)z\,,
\end{equation}
where $n$ is the refractive index of the mask substrate which is assumed wavelength-invariant at a first order. The nominal $\pi/2$-phase shift corresponds to an optical path difference of $\lambda/4$ at the design wavelength.

Second, the physical mask diameter $d$ is fixed, whereas the diffraction-limited PSF scales as $\lambda$. The relative size of the mask therefore varies as $1/\lambda$, leading to chromatic changes in the reference beam in the re-imaged pupil plane.

Both chromatism effects reduce the fringe visibility observed in the relayed pupil plane as the spectral bandwidth of work increases, lowering the sensor sensitivity. 
The increased photon flux in broadband light partially compensates for this sensitivity loss, particularly in photon-limited regimes. Recent theoretical work by \citet{Haffert2024} shows that for a amplitude aberration of 0.1\,rad, the ZWFS maintains a measurement error at the level of only 1\% for a 20\% spectral bandwidth. This result indicates that chromaticity does not constitute a fundamental limitation for the ZWFS, and provides encouraging evidence that broadband operation remains viable. Building on this, our work aims to experimentally validate this capability and assess the practical performance of the sensor in wide spectral bands. Advanced multi-wavelength reconstruction techniques can also mitigate the ZWFS chromaticity and extend its dynamic range \citep{Haffert2024,Darcis2025}.

In summary, the ZWFS provides optimal performance in monochromatic light. Extending its operation to broadband light is necessary for high-contrast instruments and requires a careful assessment of the associated chromatic limitations. Understanding these effects is essential for evaluating the performance of a ZWFS-based second stage AO loop, which constitutes the objective of the experimental study presented in the following sections.

\section{The GHOST testbed}

Hosted at ESO Garching in Germany, GHOST \citep{Engler2024_GHOST} is an AO testbed which is designed to evaluate advanced wavefront sensing and control strategies under controlled laboratory conditions. Its modular architecture allows for switching between different illumination sources, wavefront sensors, and controllers, making it a versatile platform for exploring multiple AO concepts within a single facility. This flexibility is essential for the development of future high-contrast instruments dedicated to exoplanet imaging, where increasingly sophisticated AO schemes should be validated in laboratory before on-sky deployment.

In this work, GHOST is used to investigate the concept of cascade AO, in which the residual aberrations left by a first XAO system are further corrected by a more sensitive second-stage loop driven by a ZWFS. This architecture directly targets the limitations of current XAO systems such as SPHERE, GPI, and SCExAO, whose residual wavefront errors currently restrict the detection of giant exoplanets to angular separations of roughly $200$\,mas and contrasts in the $10^{-4}$ -$10^{-6}$ range.
 By combining cascade AO with a Lyot coronagraph \citep{Lyot1932,VilasSmith1987ApOpt}, GHOST enables a quantitative assessment of the contrast gains achievable with such an approach. Its flexible design also makes it an ideal platform to reproduce a wide range of observing scenarios and to bridge the gap between laboratory demonstrations and future on-sky implementation.

\subsection{Optical layout}
\begin{figure}
  \centering
  \includegraphics[width=0.48\textwidth]{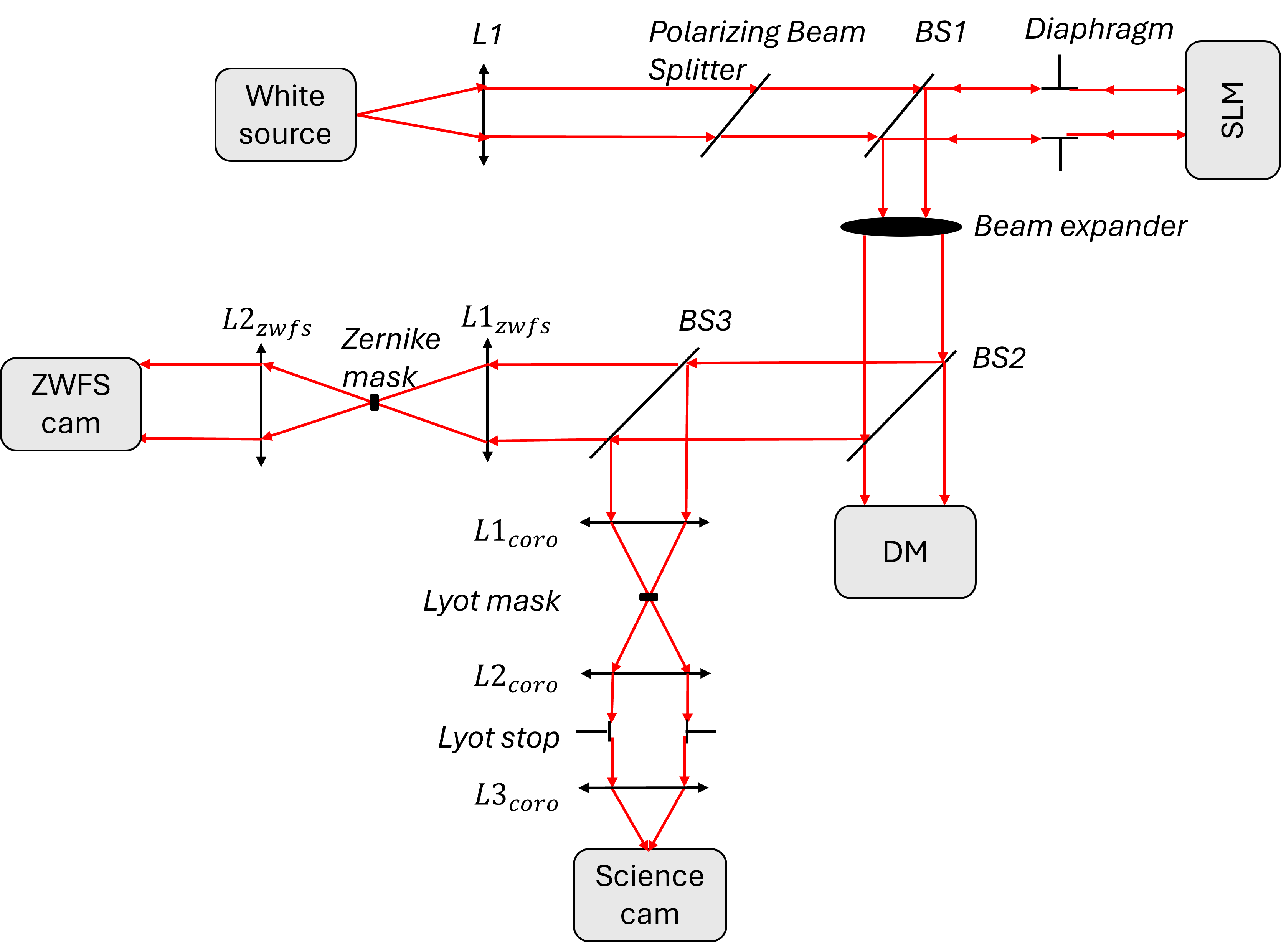}
  \caption{Optical layout of the GHOST testbed. The system includes lenses $L$, beam splitters (BS), a spatial light modulator (SLM) used to introduce XAO residuals, a deformable mirror (DM) for wavefront correction, a wavefront sensing arm incorporating the Zernike phase mask, and a science arm equipped with a Lyot mask, a Lyot stop, and a camera.}
  \label{fig:optical_layout_ghost}
\end{figure}

Figure~\ref{fig:optical_layout_ghost} presents the optical layout of the testbed. The illumination is provided with either a broadband white light lamp equipped with adjustable spectral filters or a monochromatic light source. When operating in broadband light, the relative spectral width is defined as $\Delta \lambda / \lambda$, corresponding to a 25\% spectral bandwidth centered on $\lambda = 800$\,nm. The light is injected into a single mode fibre to provide a stable diffraction-limited point source, the output of the fibre is collimated by an achromatic lens, after which a polarizing beam splitter selects the appropriate polarization state and directs it toward the spatial light modulator (SLM).
 The beam then passes through the first beam splitter BS1, which directs a fraction of the light towards a liquid crystal in silicon (LCOS) SLM.
This device introduces phase screens wich are representative of SPHERE-like residuals. They correspond to images with a wavefront error of about 90\,nm RMS in H-band under standard conditions of observations and an effective frame delay of $\sim 2$\,ms. These phase screens emulate the output of a first stage XAO system and represent the input residual wavefront errors for the second stage loop. Further details of the phase screen injection are given in \citetalias{NDiaye2024}.

After reflection on the SLM, the beam is sent back by BS1 toward a beam expander, while a portion of the light is directed via BS2 to  the Boston Micromachines 492-1.5 deformable mirror (DM) to correct for the wavefront errors. The beam is then directed toward beam splitter BS3, which divides it into two paths: the science arm and the wavefront sensing arm of the second-stage AO system. The science arm contains a Lyot coronagraph with a focal plane mask (FPM) with a relative size of 3.8\,$\lambda/D$ (at the central wavelength) and a pupil stop with a diameter of 0.84\,D. The system also includes a $256\times256$ pixel camera with a sampling of $3.99$ pixels per $\lambda/D$ at $800$\,nm. The wavefront sensing arm includes a first lens which focuses the beam onto the Zernike mask, followed by a second lens to form the relayed pupil image with a sampling of 36 pixels across the pupil diameter on the ZWFS camera. Its focal plane mask is a $40\,\mu$m dot mounted on motorised stages to ensure accurate centering. The ZWFS camera records pupil plane intensity maps used to estimate the wavefront errors. An additional field stop of 35 $\lambda/D$ (not represented in the optical layout) is also used in the ZWFS arm to minimise aliasing effects in the wavefront sensing measurements and filter optical ghosts introduced by the transmissive optics of the setup.

The optical layout uses relay lenses, introducing chromatic aberrations which have not been quantified with a direct experiment in this study to determine their impact on the ZWFS arm. However, the expected magnitude of these effects can be estimated from the standard manufacturer specifications for the achromatic doublets used in our testbed. Over the 700--900\,nm range, such lenses typically exhibit a relative focal shift of $\Delta f / f \simeq 0.1$--$0.3\%$, which corresponds to an axial defocus of $\Delta z \simeq 0.05$--$0.10$\,mm for the 100--150\,mm focal lengths optics implemented on GHOST. With a $F/50$ focal ratio beam at the level of the Zernike mask, this error translates into a defocus term of approximately $1$\,nm RMS, corresponding to a phase bias of about $0.006$--$0.012$\,rad. This chromatic error is well within the linear operating regime of the ZWFS. This amount of chromatic defocus behaves as a small, static calibration bias. A careful calibration of this effect will avoid any impact on the broadband wavefront reconstruction of the ZWFS.

\subsection{Chromaticity of the SLM}
Liquid-crystal SLMs introduce a geometric optical path difference (OPD) $\delta_{SLM}$ that depends on the effective birefringence of the liquid crystal layer. Over a limited spectral band, $\delta_{SLM}$ is nearly wavelength-independent, while the associated phase delay $\phi_{SLM}$ scales as $2\pi/\lambda$ and is therefore intrinsically chromatic. 
This effect becomes critical when the SLM is used to generate large phase excursions (typically above the $1$–$2\pi$\,rad range), where phase wrapping amplifies wavelength-dependent errors.
Indeed, the SLM produces a nearly achromatic geometric wavefront delay, meaning that $\delta_{SLM}$ (in nm) is similar at all wavelengths. The corresponding phase delay $\phi_{SLM}$, however, varies with wavelength according to 
$2\pi\,\delta_{SLM}/\lambda$ so it is chromatic.
In our configuration, the SLM is however exclusively used to reproduce low-amplitude XAO-like phase residuals, with peak-to-valley excursions far below the wrapping threshold. In this regime, the device reproduces stable and repeatable wavefront shapes across the considered spectral band.

We experimentally confirmed that the SLM exhibits consistent behaviour across the considered spectral band by injecting well-defined tip–tilt patterns and verifying their stability. Tip-tilt represents low-order, controlled aberrations that produce measurable displacements of the PSF. Any significant chromatic response would appear as wavelength-dependent variations of these displacements. In our tests, we observed no such variations. This result provides experimental evidence that the SLM gives a spectrally consistent wavefront modulation, and chromatic effects do not impact the observables relevant to this study.

\subsection{Control System and closed-loop Configuration}

The real-time control of the testbed is ensured by COSMIC \citep{Ferreira2022COSMIC}. This platform provides the real-time computation pipeline with the possible implementation of different control laws. In our configuration, the control loop is based on the use of a classical integrator.

Before closing the second loop, a calibration matrix is derived to establish the relationship between a given mode introduced on the DM and the corresponding ZWFS pupil-plane intensity map recorded on the WFS camera. These intensity patterns constitute the sensor signal used to build the interaction matrix.

First, the SLM is set to a flat position. A first interaction matrix between the DM and the ZWFS is then measured by successively applying Hadamard modes to the DM and recording the corresponding ZWFS response for each mode \citep{Kasper2004}. This interaction matrix is acquired while the ZWFS operates in the presence of aberrations on the Zernike phase mask.
From this first interaction matrix, a control matrix is computed using regularised pseudo-inversion and is used to close the control loop by imposing a zero reference on the ZWFS. The loop then compensates for the static aberrations of the optical bench, driving the DM toward a position corresponding to a flat wavefront on the Zernike mask. The system is thus brought close to the optimal operating point of the ZWFS.

A second interaction matrix is subsequently measured, using this DM position as the starting point. The same Hadamard modes are applied to the DM, but the ZWFS now operates around its zero aberration position. This second calibration provides an interaction matrix that is more representative of the actual closed-loop operating regime and improves the stability and robustness of the control. For both calibration, each column of the interaction matrix, corresponding to the WFS response to a DM excitation, is projected onto a Karhunen–Loève (KL) modal basis.
 This operation rewrites the sensor response in terms of KL modes rather than Hadamard modes, ensuring an orthogonal and turbulence optimised modal representation.
The final control matrix is computed by regularised pseudo-inversion of this projected interaction matrix and is used to convert the ZWFS measurements into modal DM commands. In this experiment, the control is limited to the first 350 KL modes, corresponding to the spatial frequency content accessible to the deformable mirror.

The temporal behaviour of the second stage loop follows the strategy established for the monochromatic light case detailed in \citetalias{NDiaye2024}. The atmospheric phase screens are generated in an  Object Oriented Python Adaptive Optic (OOPAO) simulation \citep{Heritier2023_OOPAO} at a frequency of 2 kHz and corrected by a simulated first-stage adaptive optics system operating at 1 kHz. This lower frequency arises from the fact that the simulated DM of the first stage is updated only every other frame, at half the turbulence sampling frequency. The resulting phase screens from these simulations are stored at their original 2KHz sampling rate in data cubes and then injected into GHOST using the SLM, whose maximum update frequency is 350 Hz. The replay of these residual phase screens using the SLM therefore reduces the sampling rate from 2kHz to 350Hz. Since the first stage correction was originally applied every second frame, it is effectively reproduced at 175Hz.  The second-stage adaptive optics loop operates at the same frequency as the SLM, that is 350 Hz. This configuration therefore reproduces the behaviour of a cascaded adaptive optics system in which the second stage operates at a frequency twice as high as the first stage, corresponding to the simulated case of a second stage operating at 2 kHz and a first stage operating at 1 kHz. The objective of this configuration is to reduce the temporal error between wavefront measurement and correction.

The second-stage loop is driven by the COSMIC real-time controller, which introduces a processing latency of approximately 110\,$\mu$s between the readout of the last wavefront sensor pixel and the command sent to the deformable mirror.
At each iteration, the ZWFS measures the residual phase aberration in the pupil. The integrator computes the update of the DM commands and the deformable mirror applies the corresponding correction, thereby closing the control loop. The integrator is characterised by a loop gain, which controls the strength of the correction at each iteration, and by a leak factor, which gradually damps the DM command in order to prevent long-term drift.

\subsection{Measurement Protocol in broadband light}

In \citetalias{NDiaye2024}, the system was operated with a monochromatic light source emitting at 770\,nm. In the present work, the use of a white light source allows us to evaluate the ZWFS-based control loop under realistic chromatic conditions. The measurement sequence begins by feeding the bench with a light source either in narrowband light obtained by inserting a spectral filter with a $4\,\%$ relative bandwidth or in broadband light produced by using a wider spectral filter with a relative bandwidth of approximately $25\,\%$.

Once the illumination is set, turbulence screens of different seeings and wind speeds are injected through the SLM to emulate different atmospheric conditions. For each spectral bandwidth, data are first acquired in open loop, with the DM kept flat. This configuration provides a direct measurement of the aberrations produced by the SLM and their impact on the coronagraphic images.

The loop is then closed by using the ZWFS-driven integrator to update the DM shape and new image sequences are recorded. These closed-loop sequences allow us to assess the ability of the second-stage AO loop to correct for the injected aberrations in both bandwidth configurations. Additional images are also obtained without injected turbulence to characterise the quasi-static aberrations of the bench. These errors originate from imperfections in the optical components and possible slight misalignments in the scientific arm that are not seen by the ZWFS in the sensing arm. They constitute the so-called non-common path aberrations (NCPA).

All the coronagraphic science images are normalised with the intensity peak of the non-coronagraphic images. Their radial profiles are expressed as a function of angular separation in units of $\lambda/D$. The ZWFS telemetry is recorded simultaneously, including the modal residuals derived from the KL decomposition, providing a temporal diagnostic of the correction performance and enabling a direct comparison between the narrowband and broadband light cases.

\section{Performance comparison in narrow and broadband light}
To characterise the impact of the spectral bandwidth on the performance of the ZWFS-based control loop, we compare the system response in narrowband and broadband light. These two configurations allow us to investigate the impact of photon flux and chromatic behaviour on the atmospheric residual aberrations after the wavefront error correction.

All measurements presented in this section are obtained under the same standard observing conditions, corresponding to average atmospheric turbulence with a simulated seeing of 0.7\arcsec\ and a wind speed of 10\,m.s$^{-1}$. These conditions are representative of median observing conditions on the site of 8-10\,m class telescopes and provide a consistent reference for comparing the two spectral configurations.

The residual aberrations injected by the SLM reproduce the behaviour of a first stage XAO system similar to SPHERE, whose spatial frequency cutoff is 20$\lambda/D$. This cutoff results from the actuator density and the temporal control bandwidth of such XAO systems, as detailed in \citetalias{NDiaye2024}. The second-stage deformable mirror has 22 active actuators across the pupil diameter, which sets a spatial frequency cutoff of 11\,$\lambda/D$ for the correctable wavefront errors. Two stellar flux regimes are explored. A bright star is simulated in a regime which is not limited by the source flux, while a faint star is reproduced by reducing this flux by a factor of 100, corresponding to a magnitude difference of $\Delta \mathrm{mag} = 5$.

To compare both bandwidth cases, we analyse the coronagraphic images acquired with the science camera in open and closed loop. Radial contrast profiles are computed from these images by azimuthally averaging their intensity. We then compute the ratio between the radial profiles obtained in open and closed loop. This yields the contrast gain as a function of angular separation for each test performed, which can then be plotted for quantitative comparison. 
System performance depends on the stellar flux, which sets the signal-to-noise ratio of the wavefront sensor measurements. Results are presented separately for the bright source and faint source regimes. The conditions of all our experiments in the following sections are summarised in Table~\ref{tab:table1} in the appendix.

\subsection{High-flux regime}

Figure~\ref{fig:bright-case-standard-condition} (top and middle rows) presents the coronagraphic images under a bright star flux regime for the two spectral configurations. In open loop, the coronagraph exhibits a large dark region extending up to approximately 20$\lambda/D$, which corresponds to the expected frequency cutoff of the first stage XAO system. In closed loop, a new high contrast region appears inside the XAO corrected area, extending up to about 11$\lambda/D$, which is consistent with the number of 350 corrected KL modes ($\sqrt{350/\pi}$). The contrast curves confirm this interpretation: the ZWFS-based control loop allows us to achieve a contrast down to approximately  $5\times10^{-5}$ in both spectral configurations, with the broadband case providing a slightly deeper contrast which can be attributed to the increased photon flux with the bandwidth. This behaviour is expected, as chromatic errors remain negligible when the ZWFS operates in a high-flux regime.

In the narrowband and broadband configurations, closing the ZWFS-based control loop provides a contrast gain of nearly a factor of 2 between open and closed-loop operation at angular separations larger than coronagraph mask radius. In the high flux regime, both configurations exhibit a very similar performance, which shows that chromatic effects do not play a limiting role.

\begin{figure}[!ht]
  \centering
  \includegraphics[width=0.48\textwidth]{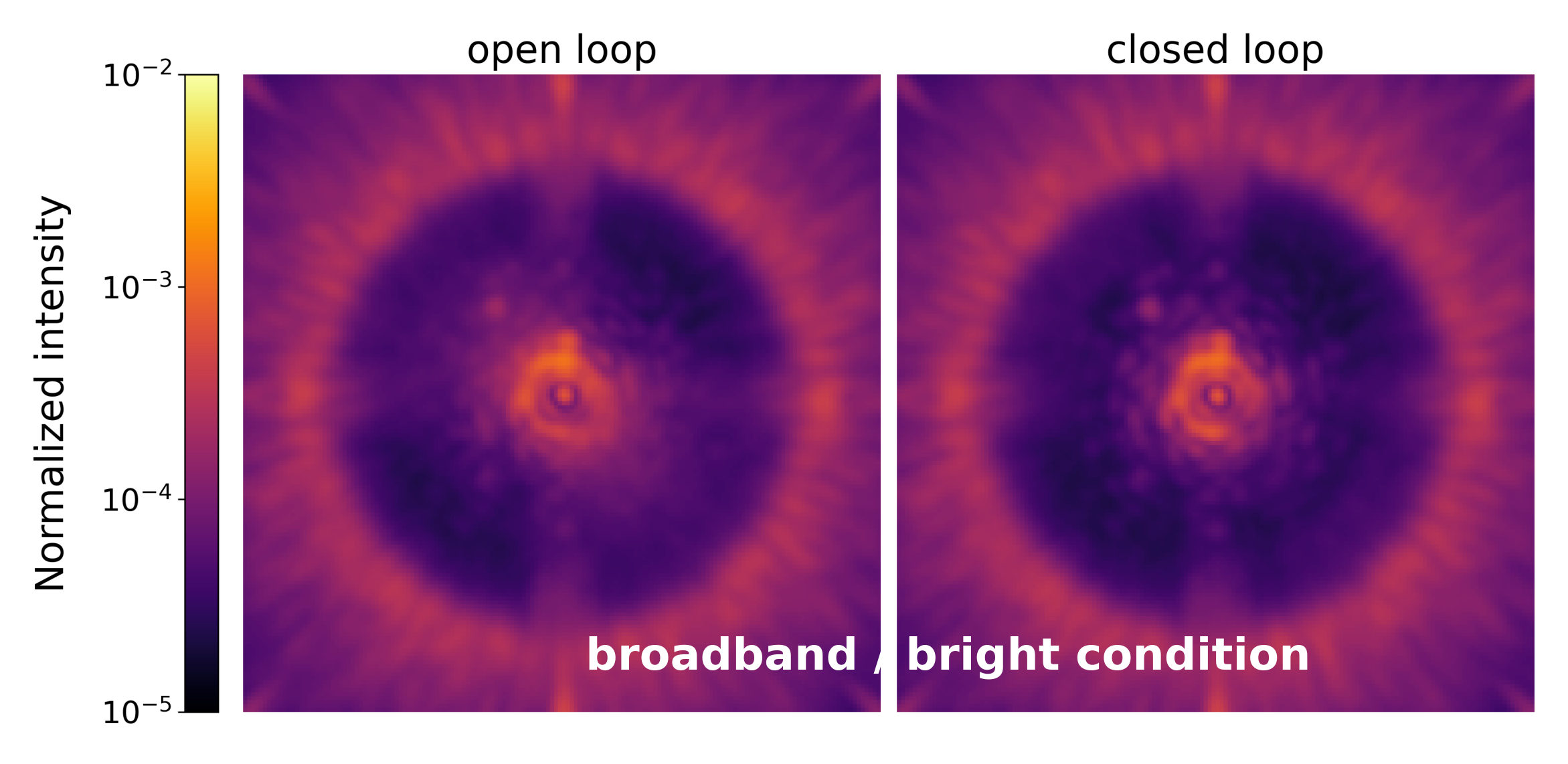}
  \includegraphics[width=0.48\textwidth]{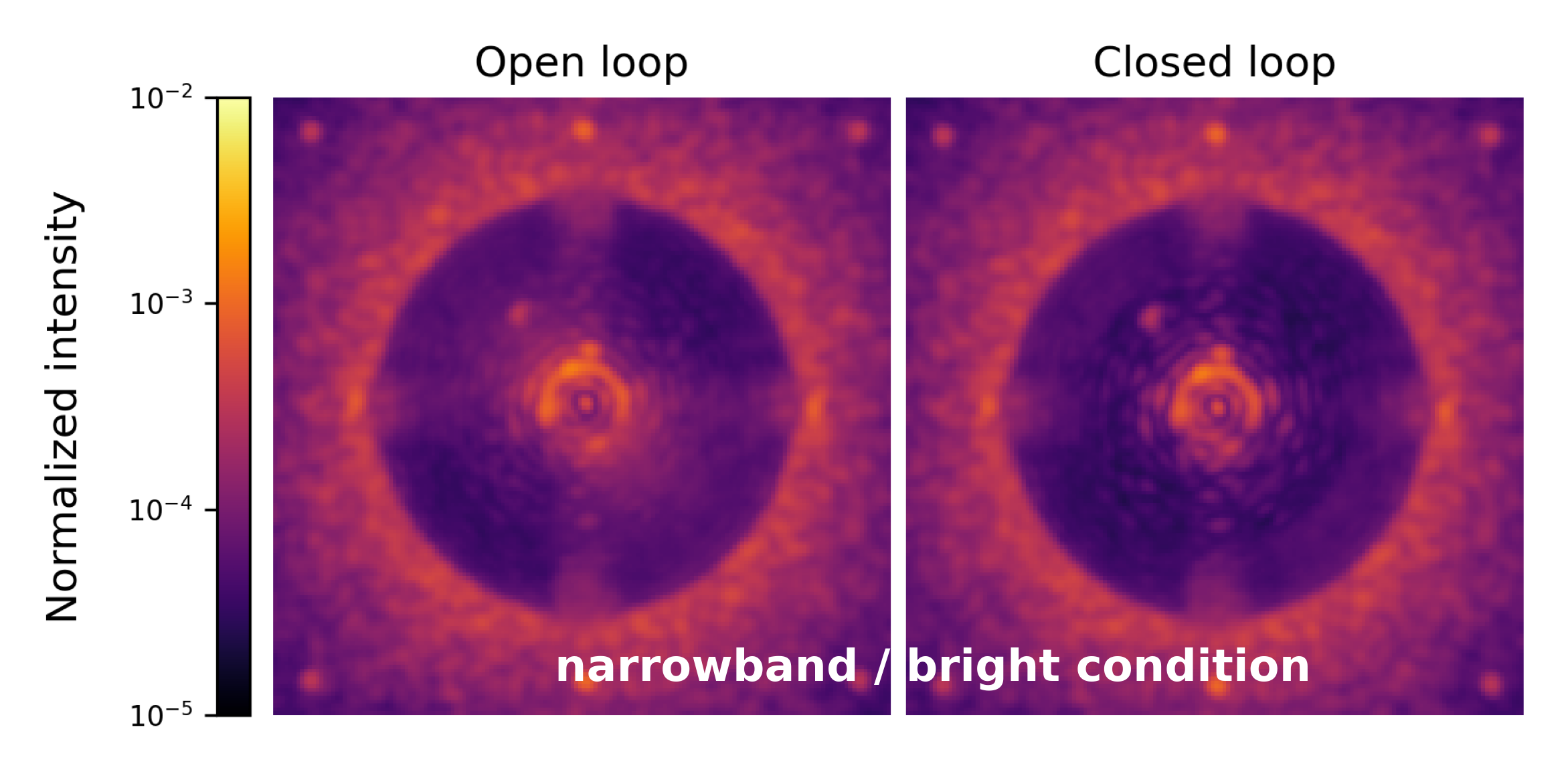}
  \includegraphics[width=0.53\textwidth]{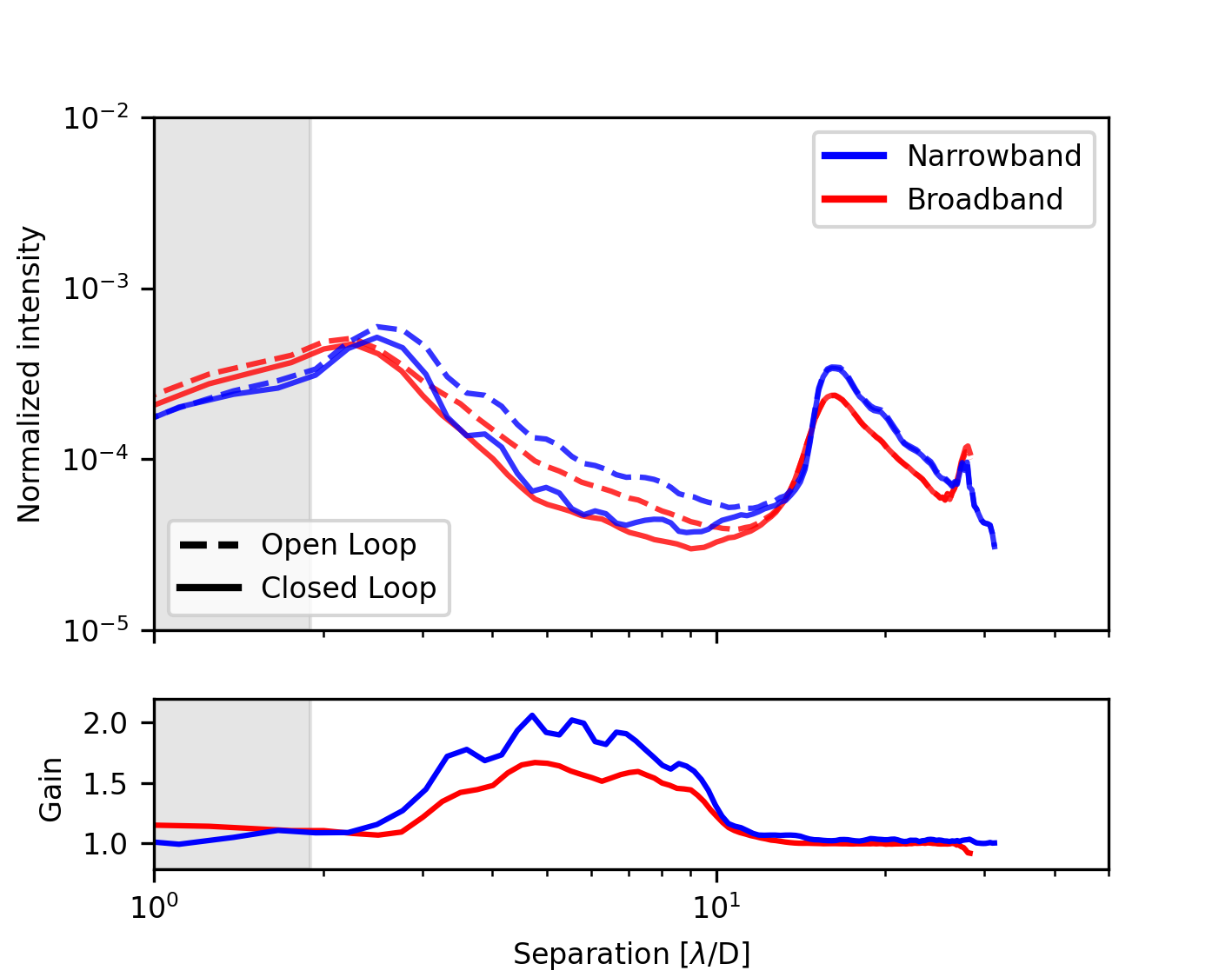}
  \caption{Comparison of contrast performance obtained in broadband and narrowband for a bright star, under standard atmospheric conditions, seeing of 0.7$^{\prime\prime}$ and wind speed of 10 m·s$^{-1}$. Top: coronagraphic images for the second stage AO in open loop (left) and closed loop (right) in broadband light. Middle: same comparison in narrowband. Bottom: contrast curves derived from the azimuthal averaged intensity of the images as a function of radial separation, with dashed lines for open loop and solid lines for closed loop. Below: gain curves computed between open and closed loop, in narrowband (blue) and broadband (red) light. The grey shaded area indicates the radius of the FPM.}
  \label{fig:bright-case-standard-condition}
\end{figure}

\subsection{Low flux regime}

\begin{figure}[!ht]
 \centering
  \includegraphics[width=0.48\textwidth]{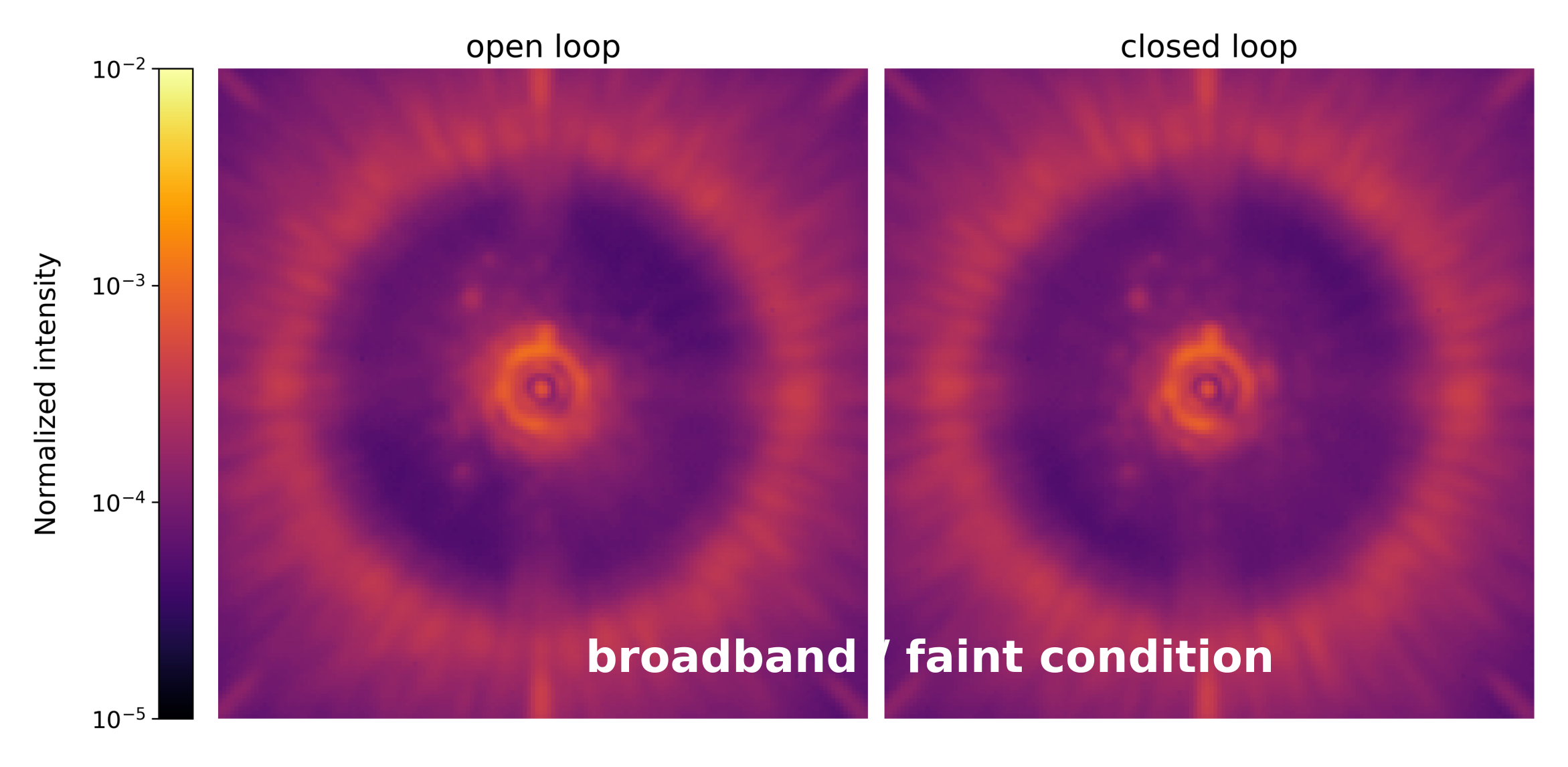}
  \includegraphics[width=0.48\textwidth]{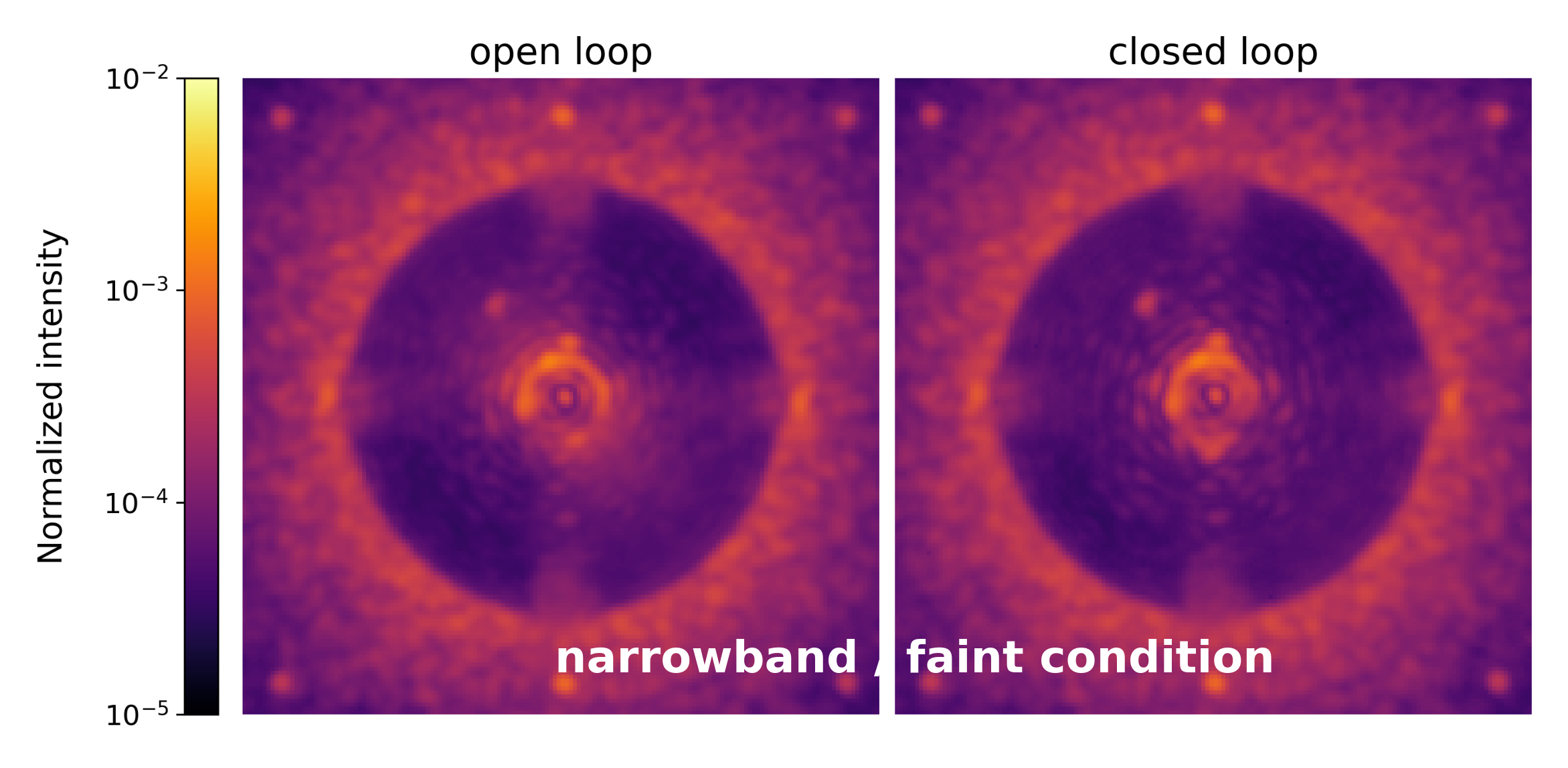}

  \includegraphics[width=0.53\textwidth]{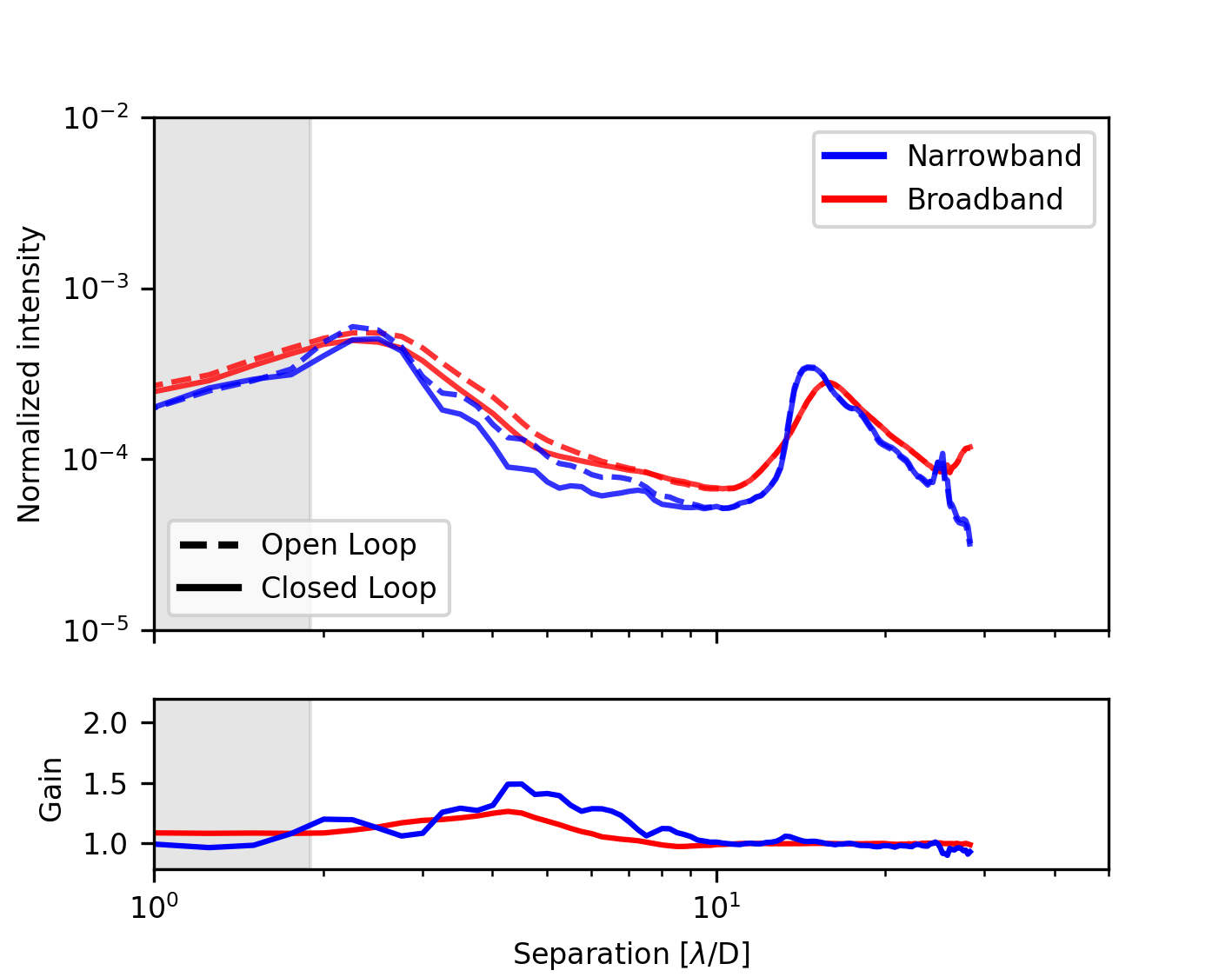}
  \caption{Comparison of contrast performance obtained in broadband  and narrowband  for a faint star, under standard atmospheric conditions (seeing of 0.7$^{\prime\prime}$ and wind speed of 10 m·s$^{-1}$). Top: coronagraphic images for the second stage AO in open loop (left) and closed loop (right) in broadband light. Middle: same comparison in narrowband. Bottom: contrast curves derived from the azimuthal averaged intensity of the images as a function of radial separation, with dashed lines for open loop and solid lines for closed loop. Below: gain curves computed between open and closed loop, in narrowband (blue) and broadband (red) light. The grey shaded area indicates the radius of the FPM.}
  \label{fig:faint-case-standard-condition}
\end{figure}

Figure~\ref{fig:faint-case-standard-condition} shows the coronagraphic images in open and closed loop and the corresponding contrast curves in the faint star configuration. 
In open loop, broadband images are naturally slightly more extended because they combine images at multiple wavelengths, even though the simulated first‑stage residuals are identical in Fig.~\ref{fig:bright-case-standard-condition} and in Fig.~\ref{fig:faint-case-standard-condition}. At high flux, this effect is hidden by the bright aberration-induced structure. At low flux, these structures are dominated by the photon noise and the broadband elongation in the images becomes less visible although the wavefront errors are unchanged between Fig.~\ref{fig:bright-case-standard-condition} and Fig.~\ref{fig:faint-case-standard-condition}.

As in the bright-star case, closing the loop improves the contrast relative to open loop. The narrowband and broadband configurations again deliver comparable performance, with both showing a slight reduction in gain compared to the bright-star case. This reduction is expected from the lower photon flux, which increases the temporal noise in the ZWFS measurement regardless of the spectral bandwidth.

Two effects govern the behaviour of the ZWFS in broadband light. The chromatic errors of the Zernike mask (wavelength-dependent phase shift and effective dot size ) reduce the intrinsic sensitivity of the sensor. However, these errors are largely static and are absorbed by the calibration matrix and the loop integrator. As a result, they do not propagate as temporal residuals and do not measurably degrade the closed-loop contrast in either flux regime. Photon noise is the main temporal error source and increases as the flux decreases. Broadband light increases the photon flux by a factor $\sim 6$ relative to narrowband, which reduces this noise. At the faintest flux level tested, this reduction is partially offset by the increased sensitivity to chromatic biases at low signal-to-noise ratio. The two effects nearly balance, so the performance of the broadband and narrowband configurations remains comparable in both flux regimes.

Overall, these results demonstrate that extending the spectral bandwidth to 25\% does not introduce a measurable degradation of the ZWFS-based control loop, even at reduced flux. The marginal advantage of narrowband at the faintest levels is not due to a fundamental limitation of broadband operation, but to the combined effect of photon noise and chromatic sensitivity at low SNR.

\subsection{Comparison with monochromatic case}
The performances obtained in this section are consistent with the monochromatic results presented in \citetalias{NDiaye2024}, under standard observing conditions. In both studies, the ZWFS-based control loop operates in a regime of small residual aberrations and provides wavefronterror reductions of the same order of magnitude. In the case of a bright star, our current tests presents exactly the same monochromatic conditions as \citetalias{NDiaye2024} and we measure a contrast gain of about a factor of two, in good agreement with the monochromatic results. In addition, previous studies reported a degradation of contrast at low flux due to the noise amplification within the controlled region, defining a flux limit for the efficiency of the second stage. This behaviour is consistent with our observations in the faint-star regime.
\section{Contrast assessment after NCPA subtraction}

The previous section validated the operation of the second control loop based on a ZWFS in broadband light. However, after correction of the dynamic residuals by the second stage, the system reaches a contrast threshold of about 5$\times 10^{-5}$ in all the configurations. This limit is thought to be related to the presence of NCPA in the testbed. This section focuses on the reduction of these quasi-static aberrations to overcome the current contrast limits and assess the performance of our control loop in the absence of NCPA.

For each spectral configuration, and for both bright and faint flux levels, coronagraphic images are recorded with and without injected turbulence, always with the Lyot mask in place and with the ZWFS second-stage in closed loop. To evaluate the intrinsic dynamical capability of the ZWFS-based control loop independently of these static limitations, we subtract the quasi-static component from the coronagraphic data. At angular separations shorter than about 3\,$\lambda/D$, the flux can locally be over-subtracted, leading to negative residuals and therefore, missing data points or apparent discontinuities in the contrast curves which are represented in logarithmic scale. These artefacts should not be interpreted as a physical limitation of our control loop.

All images are normalised to the intensity peak of the non-coronagraphic PSF and aligned with sub-pixel accuracy before subtraction. The resulting frames isolate the purely atmospheric residuals corrected by the second stage loop and reveal the true performance of the controller across its full correction domain, extending from the coronagraph edge to approximately $11\,\lambda/D$. 

\subsection{Standard atmospheric conditions}
\begin{figure}[!ht]
  \centering
  \includegraphics[width=0.48\textwidth]{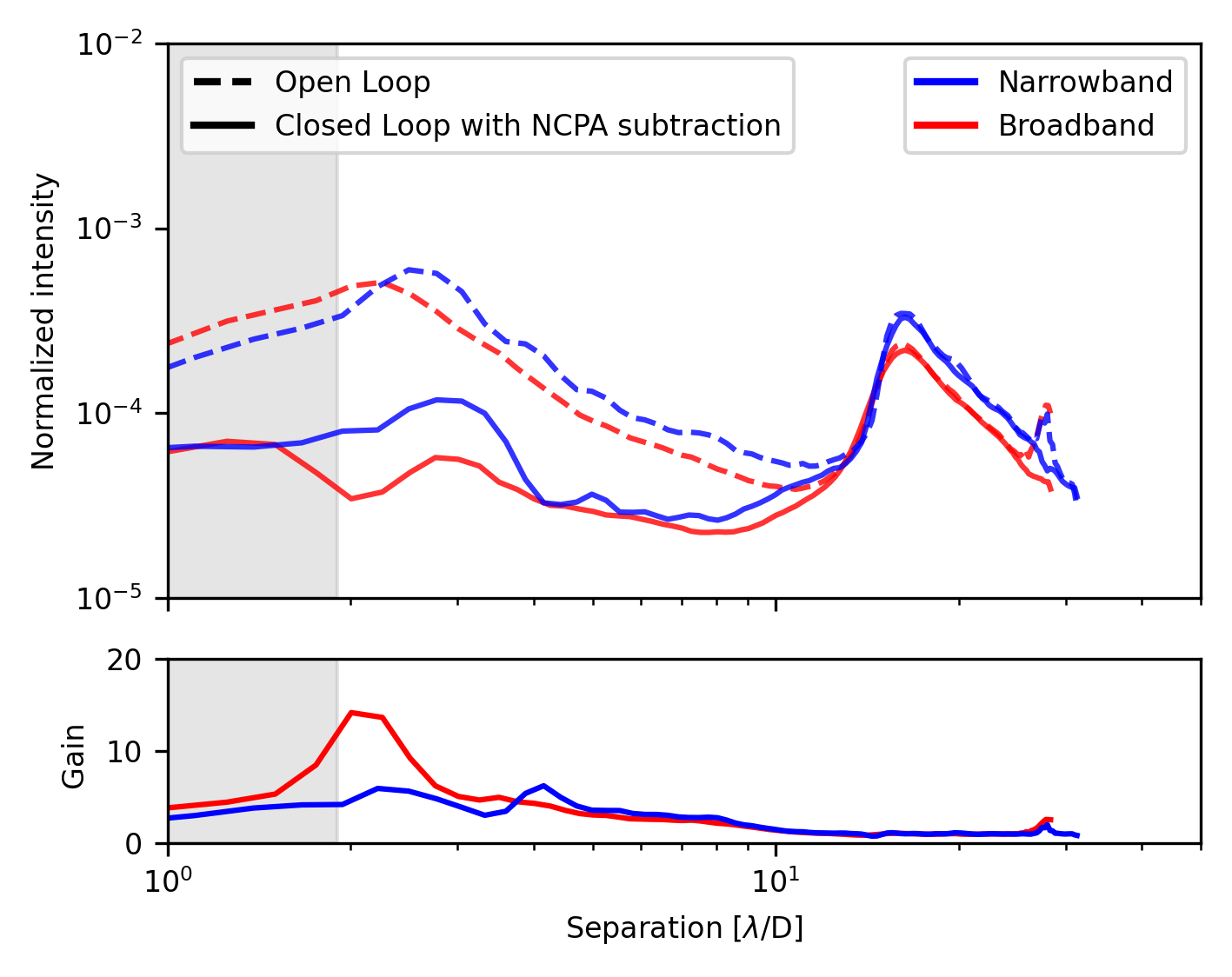}
  \caption{Contrast curves obtained after subtraction of NCPA for the bright star configuration, under standard atmospheric conditions (seeing of 0.7$^{\prime\prime}$ and wind speed of 10\,m$\cdot$s$^{-1}$). Top: azimuthally averaged coronagraphic intensity profiles in open loop (dashed) and in closed loop after NCPA subtraction (solid), shown for the narrow spectral band (blue) and the broadband configuration (red).
Bottom: corresponding gain curves, defined as the ratio between the open loop and the NCPA corrected closed loop profiles. The grey shaded area indicates the radius of the FPM.}
  \label{fig:ncpa-standard-cond}
\end{figure}

Figure~\ref{fig:ncpa-standard-cond} shows the resulting contrast curves after subtraction of the quasi-static component, for the bright star configuration under standard observing conditions ($0.7''$ seeing, 10\,m.s$^{-1}$ wind speed). 
After NCPA subtraction, the contrast curves exhibit a clear reduction of the atmospheric residuals. In narrowband and broadband alike, the closed loop contrast lies roughly one order of magnitude below the open loop level throughout the 2-11\,$\lambda/D$ region. 

The corresponding gain curves confirm a contrast improvement of about a factor of five across nearly the entire control domain.
Once the quasi-static speckles are removed, the narrowband and broadband results become very similar, confirming that the chromatic differences observed in the raw curves in Fig.~\ref{fig:bright-case-standard-condition}  were dominated by static aberrations rather than by intrinsic limitations of the ZWFS-based control loop. 
In our configuration, the purpose of using broadband illumination is not to improve the contrast relative to narrowband light, but to ensure that the increased spectral width does not degrade the performance of the ZWFS-based control loop. The chromatic errors of the ZWFS behave mainly as a static bias that is absorbed by the calibration matrix and does not propagate temporally. As a consequence, the contrasts obtained in narrowband and broadband light remain comparable.  Below 3 $\lambda/D$, the small difference between broadband and narrowband comes from the NCPA subtraction quality, not from the spectral bandwidth. The reference image subtraction can vary slightly between measurements, independent of the spectral configuration. At larger separations, the contrast is dominated by the first-stage residuals and the coronagraph response, so no improvement is expected from a wider bandwidth.At larger separations, the contrast is dominated by mid and high order spatial frequencies set by the first-stage AO residuals and by the coronagraph response, so no contrast improvement is expected with the enlargement of the spectral bandwidth.

Under standard conditions, the second stage therefore provides a robust improvement in contrast, significantly greater than what the raw contrast curves suggested in the previous sections.

\subsection{Strong turbulence conditions}
\begin{figure}[!ht]
  \centering
    \includegraphics[width=0.48\textwidth]{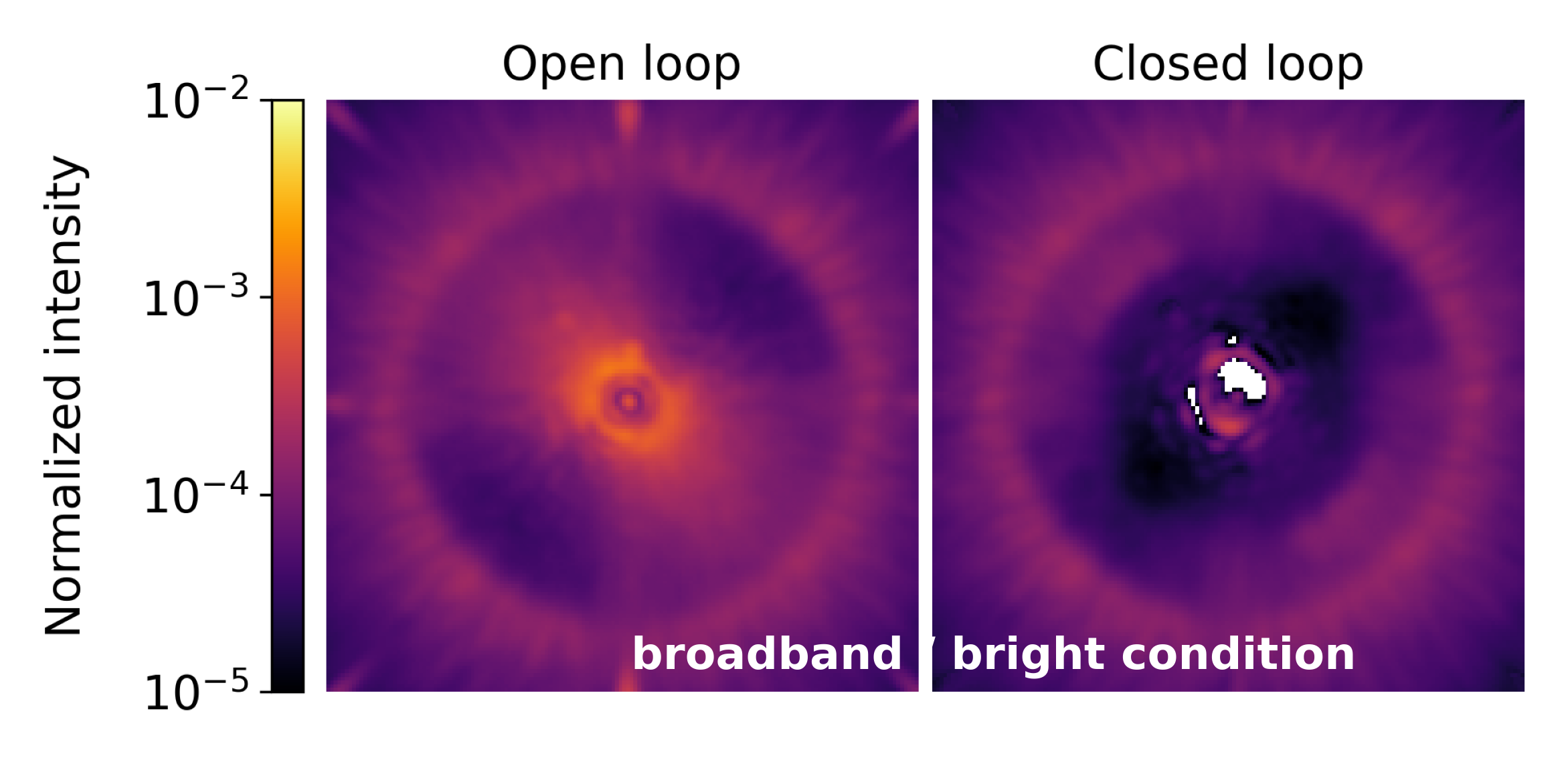}
  \includegraphics[width=0.48\textwidth]{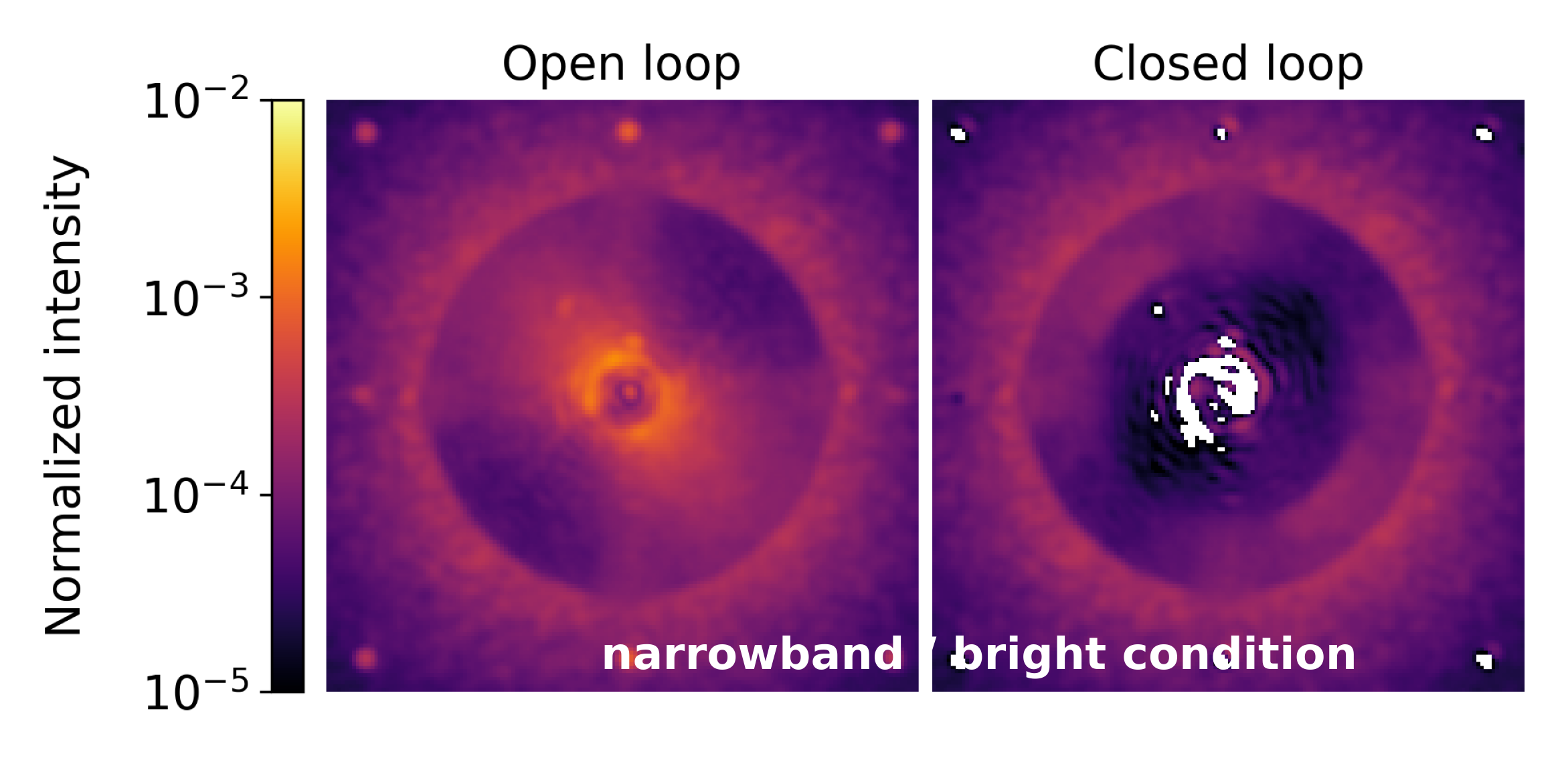}
  \includegraphics[width=0.48\textwidth]{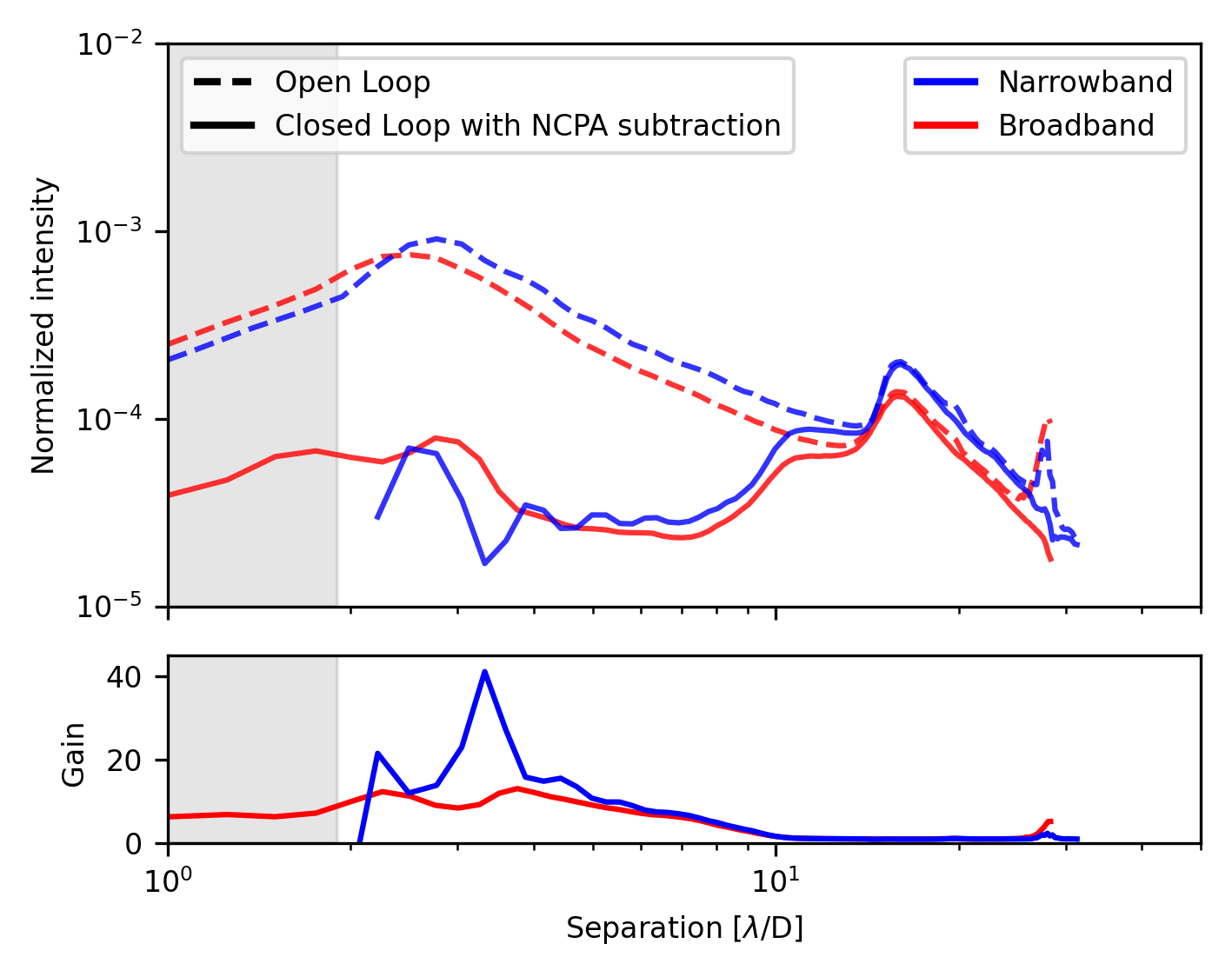}
  \caption{Contrast curves obtained after subtraction of Non Common Path Aberrations (NCPA) for the  bright star configuration, stronger atmospheric turbulence (seeing of 0.5" and wind speed of 27\,m.s$^{-1}$).Top: coronagraphic images for the second stage AO in open loop (left) and closed loop (right) in broadband light after ncpa subtraction. Middle: same comparison in narrowband.
  Bottom: open loop (dashed) and NCPA subtracted closed loop (solid) coronagraphic profiles for narrowband (blue) and broadband (red) illumination.
Below: gain curves. The grey shaded area indicates the radius of the FPM.}
  \label{fig:ncpa-strong-cond}
\end{figure}

Figure~\ref{fig:ncpa-strong-cond} presents the images in open loop and in closed loop after NCPA subtraction and the resulting contrast curves under stronger turbulence driven by the higher wind speed ($0.5''$ seeing, 27\,m.s$^{-1}$ wind speed). In this regime, the loop must operate with a substantially reduced loop gain to remain stable, and the raw contrast before subtraction of the quasi-static component is severely impacted by the rapidly evolving turbulence. 
After subtraction of the quasi-static aberrations, a significant improvement is observed: the contrast gain reaches a factor up to ten over most of the control region. With our control loop, the strong atmospheric conditions lead to a higher contrast gain than the standard atmospheric conditions. This can be explained by the fact that in this more degraded regime the open loop contrast is more strongly impacted by dynamic aberrations, thereby providing a larger correction margin for the second stage.

After subtraction of the quasi-static component, the ZWFS-based second stage loop reveals its intrinsic behaviour: within the 2-11\,$\lambda/D$ region, it consistently provides a contrast gain reaching at least one order of magnitude, independently of spectral bandwidth. This confirms that the contrast floor observed in the raw data is primarily set by the quasi-static aberrations which are not seen by the ZWFS sensing path. At the same time, the difficulty of closing the loop in very fast or strong turbulence is unrelated to NCPA and instead results from the limited dynamic range of the ZWFS itself. The following section is dedicated to testing the second control loop under different observing conditions to evaluate its limitations after reduction of the quasi‑static aberrations.

\section{Influence of observing conditions on broadband performance}
To assess the robustness of the ZWFS-based control loop under realistic observing conditions, we investigate the evolution of the performance in broadband light, for different temporal dynamics, spatial strength of the turbulence, and photon budget. These three aspects correspond respectively to changes in wind speed, seeing, and stellar flux on sky. In all our experiments, the control architecture and latency remain unchanged, and the loop always operates on the same set of 350 KL modes, corresponding to a correction radius of about 11\,$\lambda/D$. For each atmospheric condition, the loop gain is adjusted to maintain stability while providing the best possible correction, as is typically done on astronomical instruments. We remind that the conditions of all our experiments are summarised in Table~\ref{tab:table1} in the appendix. The closed‑loop results are here presented after NCPA reduction, following the method detailed in the previous section. To ensure an optimal correction of the quasi-static errors, the reference images without turbulence are scaled to compensate for differences in exposure time and illumination level with the acquired coronagraphic images for a given configuration.

\subsection{Effect of wind speed}
Wind speed governs the rate at which turbulent structures sweep across the telescope pupil and therefore sets the temporal frequencies present in the residual aberrations. It therefore plays a key role in the performance of the XAO system and in the residuals transmitted to the second stage. When the wind is fast, these structures evolve rapidly, producing turbulence with high temporal frequencies which challenges the ability of a simple integrator to follow the wavefront evolution. The resulting temporal phase error $\sigma_{t}$ scales approximately as
\begin{equation}
\sigma_{t}^{2} \simeq (2\pi f_{t}\,\tau)^{2}\,,
\label{eq:temporal_error}
\end{equation}
where $f_{t}$ denotes the characteristic temporal frequency of the turbulence and $\tau$ is the effective frame delay of the second-stage loop. \citep{Fried1990,Conan1995}

When the wind speed is low, the residuals evolve slowly and the second stage loop can provide significant correction even with a relatively high gain. As the wind speed increases, the turbulence enters a regime in which the integrator can no longer keep up with the temporal evolution, and the loop has to be stabilised by reducing the control loop gain. This constraint inevitably reduces the correction efficiency at high wind speeds.

Figure~\ref{fig:windspeed} presents the contrast curves obtained under various observing conditions, highlighting the residual wavefront errors transmitted by the XAO first stage at different wind speeds. {At moderate and high wind speeds (10 and 24\,m.s$^{-1}$), the loop operates efficiently and provides a significant contrast improvement by a factor of nearly 10, respectively, inside the control radius. For a wind speed of 5\,m.s$^{-1}$, the turbulence evolves slowly, which reduces the amplitude of the residuals transmitted to the second loop. The contrast gain measured between open and closed loop therefore appears less pronounced than in the previous cases since the first loop already corrects for the majority of the wavefront errors. Even with a high loop gain of 0.8, the second loop has very small aberrations left from the first-stage XAO to compensate for.
At very high wind speeds (34\,m.s$^{-1}$), the turbulence becomes too fast to follow, and the loop only remains stable with a very small loop gain of 0.03, leading to nearly identical contrasts in open and close loops.

These results reflect the expected temporal behaviour of atmospheric turbulence and show that the temporal bandwidth of the second stage loop is fundamentally limited by its effective correction frequency and by the ability of the control algorithm to cope with rapidly evolving conditions. When the turbulence becomes too fast, the loop gain has to be be reduced drastically to avoid divergence, which restricts the ability of the second stage to correct for the residuals from the first XAO stage. A possible strategy to alleviate this limitation is to reduce the number of corrected modes, enabling higher loop gains and improving temporal robustness, as demonstrated in \citetalias{NDiaye2024}. While this optimisation was not implemented in the present work, it may offer a possible avenue for improving the stability and overall efficiency of a ZWFS-based control loop under fast evolving turbulence. However, its effectiveness is not straightforward and would require further investigation, which is out of the scope of this paper.

\begin{figure}[!ht]
  \centering
  \includegraphics[width=0.48\textwidth]{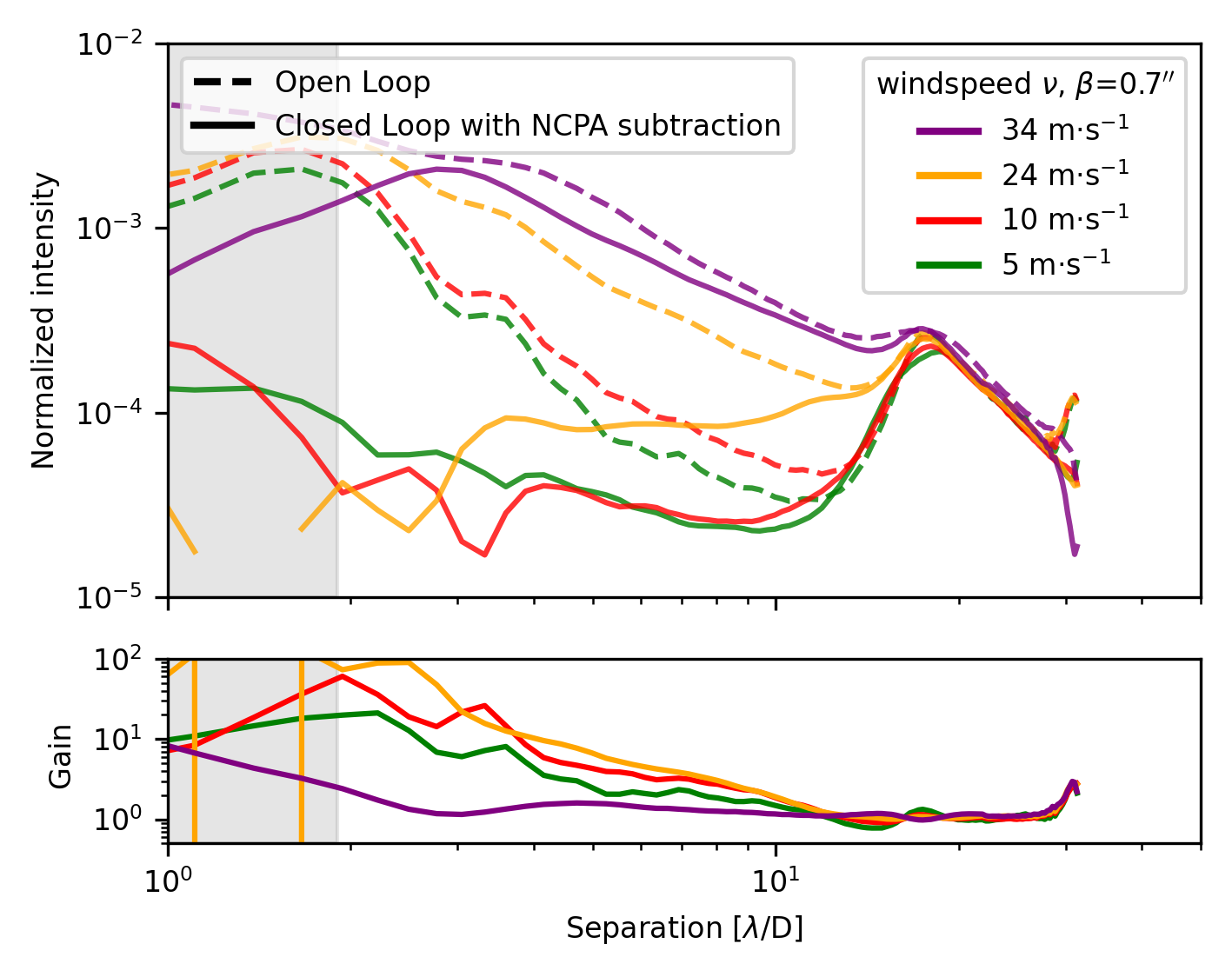}
  \caption{Contrast curves of the coronagraphic images obtained in broadband light and in bright star conditions for different wind speeds, with a seeing of 0.7$^{\prime\prime}$. Top: azimuthal averages of the images in open loop (dashed lines) and closed loop (solid lines), plotted as a function of radial separation (in $\lambda/D$) for the different wind conditions. Bottom: gain  curves in log scales computed between the open loop and closed loop cases for the various wind speeds. The grey shaded area indicates the radius of the FPM.}
  \label{fig:windspeed}
\end{figure}

\subsection{Effect of seeing}
The seeing $\theta_{\mathrm{seeing}}$ reflects the atmospheric conditions during observation. Its value has an impact on the performance of the first XAO stage and, therefore, on the residuals which are transmitted to the second-stage AO. A poor seeing corresponds to a more distorted incoming wavefront, while a good seeing indicates weaker phase fluctuations. This behaviour is quantified through the Fried parameter $r_0$, which measures the spatial coherence length of the turbulence \citep{Fried1966,Fried1966b,Sarazin1990}. The relationship between seeing and $r_0$ writes as 
\begin{equation}
\theta_{\mathrm{seeing}} \simeq 0.98 \frac{\lambda}{r_0}\,,
\label{eq:seeing_r0}
\end{equation}
so that a degraded seeing corresponds to a smaller $r_0$. This reduction in $r_0$ leads to a rapid increase in turbulence strength, because the phase variance $\sigma_{\phi}$ over a pupil of diameter \(D\) scales as  
\begin{equation}
\sigma_{\phi}^{2} \propto \left( \frac{D}{r_0} \right)^{5/3}\,.
\label{eq:phase_variance}
\end{equation}

A poor seeing (small \(r_0\)) therefore implies a larger phase variance, meaning that the first stage AO system leaves stronger residual aberrations to be corrected by the second stage \citep{Fried1966,Fried1966b}.
As these residual distortions increase, the ZWFS progressively moves away from its optimal small aberration regime. Its quasi-linear operating range is valid for phase errors below a few tenths of a radian RMS equivalent to a few tens of nanometers RMS in the visible on the GHOST testbed. When the aberration amplitude exceeds this capture range, the non-linearity of the sensor response leads to a degradation of the phase reconstruction accuracy. In practice, the control loop is forced to operate with a lower gain to remain stable, which inevitably limits the correction efficiency.

Figure~\ref{fig:seeing} shows the contrast curves for different seeing conditions, illustrating the residuals left by the first-stage correction that are subsequently handled by the second stage. For seeing values of 0.5'' and 0.7'', the loop can be operated with high gains and provides a clear contrast gain within the 11 $\lambda/D$ correction radius. When the seeing degrades to 1.0'', the injected phase variance increases substantially and pushes the second stage closer to the limit of its linear operating range. In this regime, the loop can only remain stable with a reduced gain of about 0.3, which naturally limits the correction efficiency. Nevertheless, the closed loop contrast curves still show a clear improvement compared to the open loop case, and the gain profile confirms that the loop continues to remove a significant amount of residual energy despite the harsher conditions. 

This result is particularly encouraging: even when the turbulence amplitude is large and the ZWFS operates near the edge of its linear range, the second stage keeps on providing a measurable contrast gain. This behaviour illustrates the value of a cascade AO architecture, where a dedicated second-stage correction of residuals aberrations improves performance beyond the limits of a single high-order XAO.
 
\begin{figure}[!ht]
  \centering
  \includegraphics[width=0.48\textwidth]{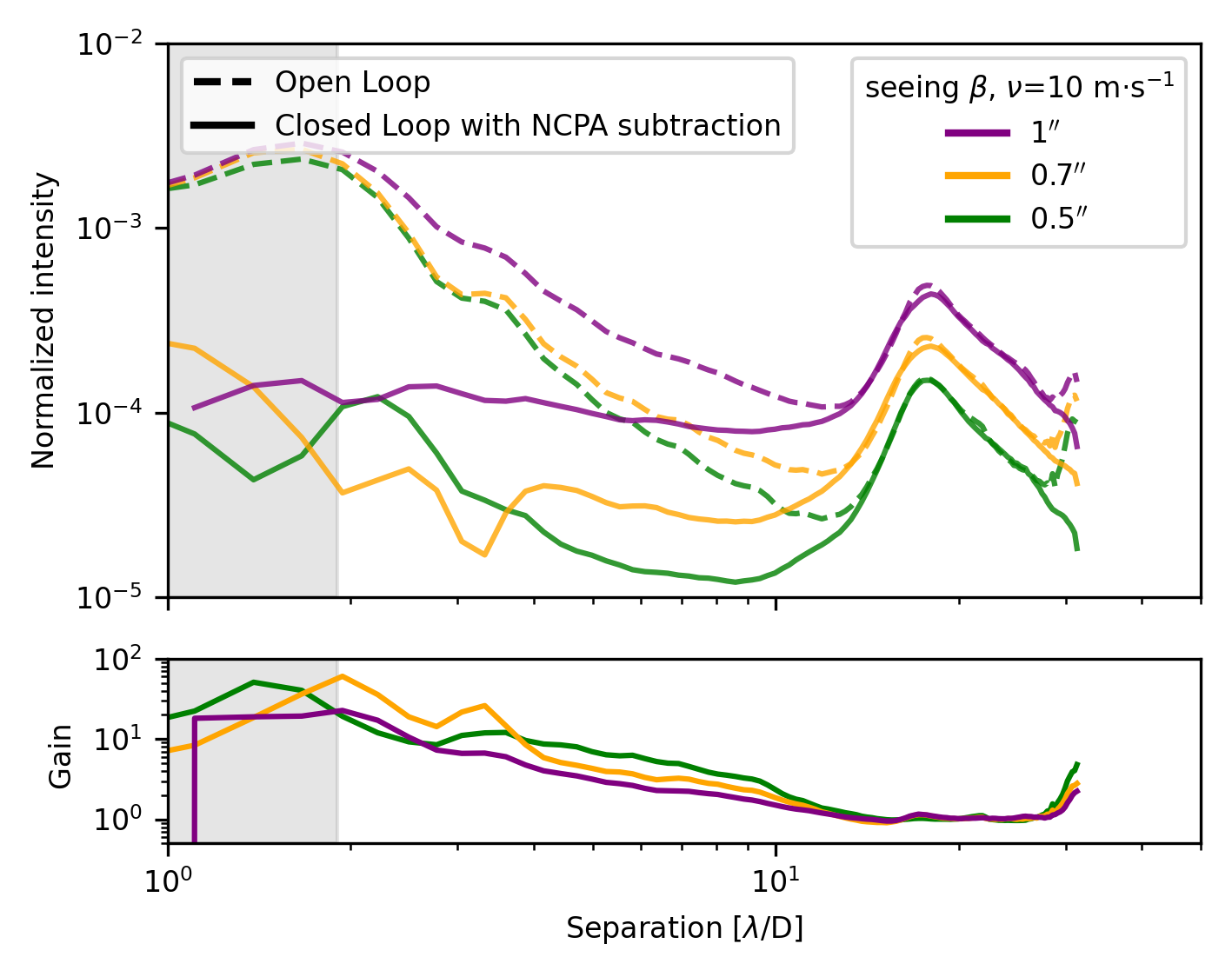}
  \caption{Contrast curves  of the coronagraphic images obtained in broadband light and in bright star conditions for different seeing conditions, with a wind speed of 10\,m$\cdot$s$^{-1}$. Top: azimuthal averages of the images in open loop (dashed lines) and closed loop (solid lines), plotted as a function of radial separation for each seeing value. Bottom: gain curves in log scales computed between the open loop and closed loop cases for the different seeing conditions. The grey shaded area indicates the radius of the FPM.}
  \label{fig:seeing}
\end{figure}

\subsection{Effect of stellar flux}
The stellar flux directly sets the signal-to-noise ratio (SNR) of the ZWFS, thereby determining whether the control loop operates in a photon-rich or photon-limited regime.
In real observing conditions, this flux is directly related to the apparent magnitude of the target star. In particular, a decrease in flux leads to a degradation of the SNR, which can limit the measurement and consequently, the correction efficiency. To explore the impact of these SNR variations on the system performance, the results are presented for different injected flux levels. The injected flux is expressed as a function of a reference level $F_{\mathrm{ref}}$, defined as 100\% of the input flux. For a given flux $F$, the corresponding magnitude difference is expressed using the standard relation
\begin{equation}
\Delta \mathrm{mag} = -2.5 \log_{10}\left(\frac{F}{F_{\mathrm{ref}}}\right) \,.
\label{eq:deltamag}
\end{equation}

In this study, three flux levels are explored. The first corresponds to the reference flux ($\Delta \mathrm{mag}=0$), the second reduces the flux to 20\% of the reference ($\Delta \mathrm{mag} \simeq 1.75$), and the third simulates a very faint source by reducing the flux to 1\% of the reference ($\Delta \mathrm{mag} \simeq 5$). These values define three observational regimes while keeping turbulence, alignment, and control parameters identical across the tests.

From a physical standpoint, the performance of the ZWFS-based control loop is governed by the balance between photon noise and chromaticity. At high-flux regime, the dominant limitation is the photon noise, and the RMS error $\sigma_\phi$ on the reconstructed wavefront phase decreases as the photon number $N$ increases, following
\begin{equation}
\sigma_{\phi} \propto \frac{1}{\sqrt{N}} \,.
\label{eq:photonnoise}
\end{equation}
In this regime, the ZWFS fully benefits from its optimal high aberration sensitivity, and chromatic effects arising from both the wavelength dependence of the mask phase shift and the scaling of the dot size in units of $\lambda/D$ remain negligible. As the source flux decreases, the signal-to-noise ratio drops and photon noise increases until it becomes comparable to the static chromatic biases, leading to a degradation of the broadband phase estimation

Figure~\ref{fig:stellar-flux} illustrates the closed loop coronagraphic images and contrast curves for the three flux levels. For the brightest case, provides a slightly deeper contrast, consistent with the improved signal-to-noise ratio, at 5$\lambda/D$, the gain reaches a factor of 4.3. In contrast, for the intermediate and faint flux levels, the performance becomes very similar, photon noise begins to limit the correction, and the contrast gain remains comparable in both cases, with a value of about 1.5 at 5$\lambda/D$.

These results show that broadband light improves performance only when the number of detected photons is sufficiently high to compensate for chromatic errors. For faint sources, photon noise amplifies the chromatic mismatch of the ZWFS, making a narrow spectral bandwidth more favourable for accurate phase reconstruction and efficient second-stage correction.

\begin{figure}[!ht]
  \centering
  \includegraphics[width=0.48\textwidth]{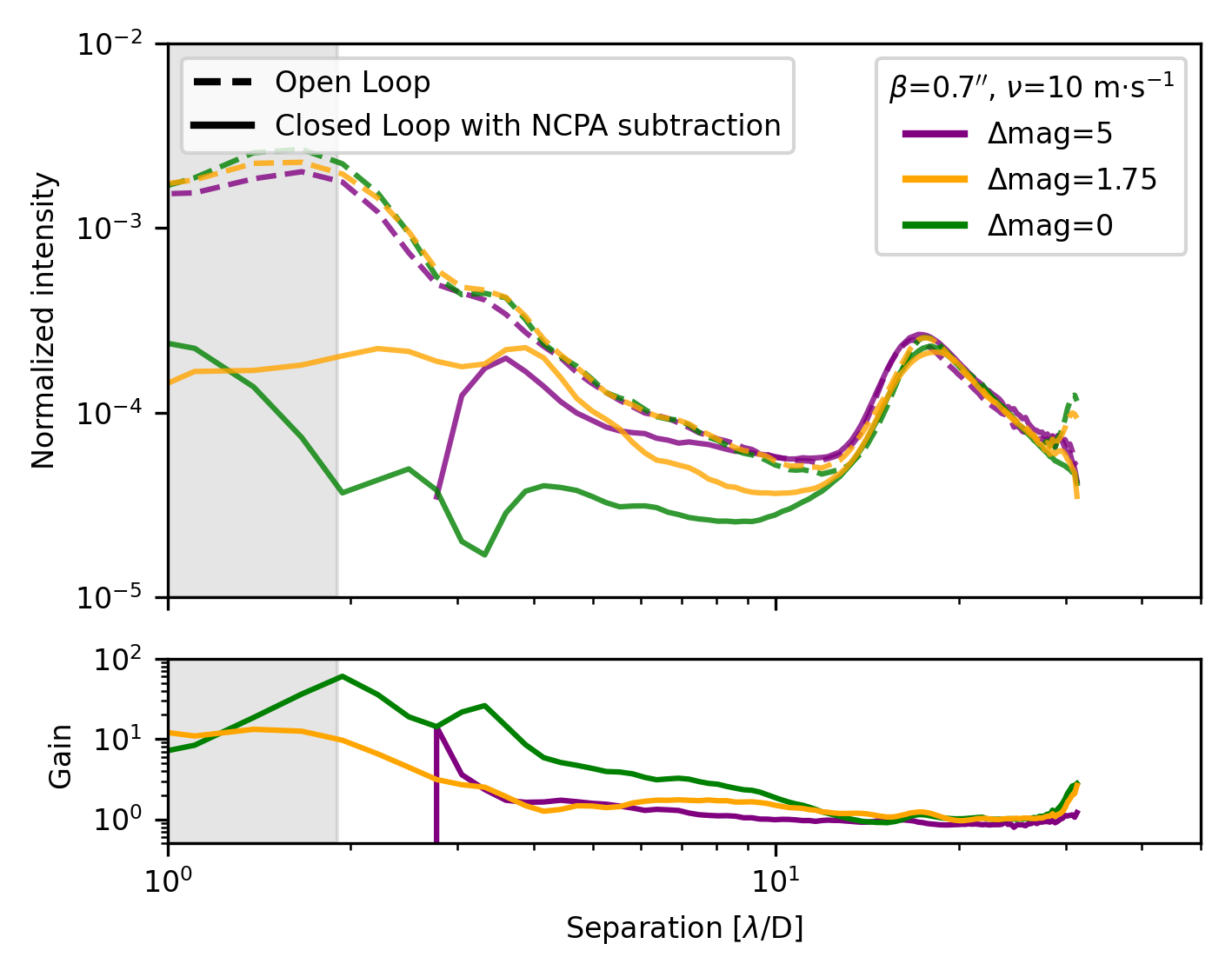}
  \caption{Contrast curves  of the coronagraphic images obtained in broadband light for different flux levels, under standard atmospheric conditions (seeing of 0.7$^{\prime\prime}$ and wind speed of 10\,m$\cdot$s$^{-1}$). Top: azimuthal averages of the images in open loop (dashed lines) and closed loop (solid lines), plotted as a function of radial separation (in $\lambda/D$) for different fluxes expressed in relative magnitude. Bottom: gain curves in log scales computed between the open loop and closed loop cases for the various flux levels.The grey shaded area indicates the radius of the FPM.}
  \label{fig:stellar-flux}
\end{figure}

\subsection{Discussion on broadband performance }

These results allow us to identify the conditions at which the ZWFS-based control loop performs optimally. Moderate wind speeds, stable seeing and high stellar flux allow the loop to operate with a high gain and to remain within the linear regime of the ZWFS, defining the most favourable operational window of the system. Under more challenging atmospheric conditions, the performance of the second-stage loop becomes limited. While quasi-static errors have already been reduced through NCPA correction, the loop cannot fully compensate for the residual turbulence when the wind speed is high, the seeing is poor or the stellar flux is low. In these cases, the standard integrator controller fails to maintain stable closed-loop operation. Extending the linear range of the ZWFS is essential to ensure stable closed-loop operation under stringent atmospheric conditions. 

%


\section{Discussion and conclusion}
In this work, we investigated the performance of a ZWFS as the basis of a second-stage adaptive optics loop, under a variety of experimental conditions. Building on previous results obtained in monochromatic light, we provide here a coherent and comprehensive validation of the ZWFS-based control loop in both narrowband and broadband light, across a wide range of turbulence strengths, wind speeds, and stellar fluxes.
Our experiments demonstrate that the ZWFS-based second stage delivers a significant improvement in contrast within its 2–11~$\lambda/D$ correction region. After subtraction of the quasi-static NCPA, the intrinsic dynamical gain delivered by the second stage reaches approximately one order of magnitude, confirming the capability of our scheme to refine the residuals left by an upstream XAO system.

Despite these encouraging results, several limitations of the current implementation become apparent. The first limitation arises from the chromatic behaviour of the scalar Zernike mask, whose wavelength dependent phase shift and effective dot size degrade the performance in broadband light, particularly for faint sources where photon noise amplifies chromatic biases. The second limitation is the presence of slow-evolving NCPA between the sensing and science channels, which must be subtracted to reveal the true dynamical performance of the loop. A third limitation concerns the classical integrator used as the temporal controller: in fast or strong turbulence, the ZWFS leaves its linear range, the loop must be operated with a very low gain to remain stable, and the correction becomes largely ineffective.

These limitations naturally point to several avenues for improvement. 
A first direction concerns the temporal controller. In addition to the classical integrator, we also tested a reinforcement‑learning‑based predictive controller  Policy Optimization for Adaptive Optics(PO4AO), recently introduced by \citet {Nousiainen2022,Nousiainen2024} under controlled conditions. These exploratory tests show that PO4AO can be stably integrated into the ZWFS-driven second stage and achieves a level of performance comparable to the integrator, with indications of improved suppression of slow temporal components. While the benefit remains modest in the narrowband, extending these tests to broadband light, stronger turbulence, or higher wind speeds will allow for a more precise evaluation of the actual performance of a ZWFS-driven second stage AO loop based on predictive control. 
A second improvement concerns modal optimisation: reducing the number of corrected modes in stringent regimes would allow higher gain and better temporal robustness, as suggested in \citetalias{NDiaye2024}. A third direction involves the correction for the NCPA, ideally in real time rather than by post-processing subtraction, for example via dedicated calibration routines or combination of measurements from multiple sensors \citep[e.g.,][]{Vigan2019}.

Beyond these improvements, several technological perspectives emerge. Recently proposed vector-Zernike masks offer an intrinsically achromatic implementation of the ZWFS phase shift by replacing the conventional scalar mask with a geometric phase half-wave mask that imposes opposite phase shifts on the two circular polarizations, allowing the simultaneous reconstruction of the amplitude and phase of the incident field and potentially alleviating the chromatic limitations observed in this work, particularly in broadband light operation \citep{Doelman2019}. Multi-wavelength ZWFS concepts provide an alternative route to extending the usable bandwidth while preserving the small-aberration sensitivity of the sensor, using dispersive or multiplexed focal plane masks \citep{Haffert2024, Darcis2025}. These developments are highly relevant for future high-contrast instruments, where broadband performance is essential to maximise the photon sensitivity and ease the spectroscopic characterisation of exoplanets.

Finally, an essential next step is the on-sky validation of this concept. A joint effort is currently underway to implement a second-stage AO loop equipped with a vector-Zernike mask on the Provence Adaptive optics PYramid Run System (PAPYRUS) platform \citep{Muslimov2021PAPYRUS,Cisse2023PAPYRUS}, a high-contrast demonstrator designed for rapid prototyping and on-sky testing of innovative wavefront control concepts. On-sky tests are planned in the near future, and will provide a decisive assessment of the maturity and practical viability of this technology under real observing conditions.

More broadly, this work strengthens the long-term perspective of integrating a Zernike-based second stage AO system into operational instruments. The first natural application is SAXO+, the planned upgrade of SAXO, the XAO system of SPHERE, where the ability of a ZWFS to reduce XAO residuals is expected to translate directly into deeper coronagraphic contrast and enhanced exoplanet detectability \citep{Boccaletti2022a}. Looking further ahead, the Planetary Camera and Spectrograph (PCS) for the ELT aims to detect temperate rocky exoplanets around nearby stars. Such a goal requires aggressive suppression of residual aberrations at separations of only a few~$\lambda/D$ \citep{Kasper2021}. This is precisely the regime in which the ZWFS-based cascade architecture is most promising.

The results of this study complete the experimental validation initiated in monochromatic light and confirm that a ZWFS-based second stage can deliver a significant and repeatable improvement of the contrast in coronagraphic images. With forthcoming developments in vector-Zernike masks, predictive control laws, and on-sky demonstrations, Zernike-based second stage adaptive optics stands out as a strong and realistic pathway for the next generation of high-contrast imagers both on current 8-10~m class telescopes and on forthcoming ELTs. 
%
%

\begin{acknowledgements} 
This project has received financial support from the CNRS through the MITI interdisciplinary programs. This work was supported by the Action Spécifique Haute Résolution Angulaire (ASHRA) of CNRS/INSU co-funded by CNES. MN acknowledges support from Observatoire de la Côte d'Azur and Laboratoire Lagrange through the 2022 BQR OCA and 2023 BQR Lagrange programs for the manufacturing of Zernike phase masks and the missions to ESO Garching. AV acknowledges funding from the European Research Council (ERC) under the European Union’s Horizon 2020 research and innovation programme, grant agreement No. 757561 (HiRISE).
\end{acknowledgements}

\bibliographystyle{aa}
\bibliography{references_ZWFS_full}

\begin{appendix}
\onecolumn
\begin{landscape}

\section{Summary of the experiments}

Table \ref{tab:table1}presents the second-stage AO loop parameters with the corresponding coronagraphic performance for all experimental configurations discussed in this paper.
\begin{table}[H]
\caption{Second-stage AO loop parameters and corresponding coronagraphic performance. }


\begin{tabular}{ccccccc|rrr|rrr|c}

\hline\hline
Bandwidth & \begin{tabular}[c]{@{}c@{}}Source \\ flux\\ $\Delta$mag\end{tabular} & \begin{tabular}[c]{@{}c@{}}Seeing\\ $\beta$\\ ''\end{tabular} & \begin{tabular}[c]{@{}c@{}}Wind\\ speed\\ $\mathrm{m\,s^{-1}}$\end{tabular} & \begin{tabular}[c]{@{}c@{}}Corr.\\ modes\end{tabular} & \begin{tabular}[c]{@{}c@{}}Loop\\ gain\end{tabular} & \multicolumn{1}{l|}{\begin{tabular}[c]{@{}l@{}}NCPA\\ subtraction\end{tabular}} & \begin{tabular}[c]{@{}r@{}}OL contrast \\  5$\lambda/D$\\ ($\times 10^{-5}$)\end{tabular} & \begin{tabular}[c]{@{}r@{}}CL contrast \\ 5 $\lambda/D$\\ ($\times 10^{-5}$)\end{tabular} & \begin{tabular}[c]{@{}r@{}}Gain  \\ 5$\lambda/D$\\ ($\times 10^{-5}$)\end{tabular} & \begin{tabular}[c]{@{}r@{}}OL contrast \\  10 $\lambda/D$\\ ($\times 10^{-5}$)\end{tabular} & \begin{tabular}[c]{@{}r@{}}CL contrast  \\ 10$\lambda/D$\\ ($\times 10^{-5}$)\end{tabular} & \begin{tabular}[c]{@{}r@{}}Gain  \\ 10$\lambda/D$\\ ($\times 10^{-5}$)\end{tabular} & Fig \\ \hline
Broadband & 0 & 0.7 & 10 & 350 & 0.8 & no & $9.64 \pm1.18$ & $5.82\pm0.71$ & $1.66\pm0.01$ & $4.02\pm0.07$ & $3.21\pm0.14$ & $1.25\pm0.07$ & $\ref{fig:bright-case-standard-condition}$ \\
Narrowband & 0 & 0.7 & 10 & 350 & 0.8 & no & $12.85\pm0.71$ & $6.56\pm0.25$ & $1.96\pm0.09$ & $5.49\pm0.11$ & $4.15\pm0.25$ & $1.33\pm0.11$ & $\ref{fig:bright-case-standard-condition}$ \\
Broadband & 0 & 0.7 & 10 & 350 & 0.8 & yes & $ 9.64 \pm1.18$ & $2.98\pm0.14$ & $3.23\pm0.25$ & $4.02\pm0.07$ & $2.73\pm0.15$ & $1.48\pm0.11$ & $\ref{fig:ncpa-standard-cond}$ \\
Narrowband & 0 & 0.7 & 10 & 350 & 0.8 & yes & $12.85\pm0.71$ & $3.44\pm0.18$ & $3.74\pm0.27$ & $5.49\pm0.11$ & $3.62\pm0.21$ & $1.52\pm0.12$ & $\ref{fig:ncpa-standard-cond}$ \\
Broadband & 5 & 0.7 & 10 & 350 & 0.4 & no & $13.88\pm1.94$ & $11.55\pm1.21$ & $1.20\pm0.04$ & $6.70\pm0.05$ & $6.81\pm0.07$ & $0.98\pm0.00$ & $\ref{fig:faint-case-standard-condition}$ \\
Narrowband & 5.40 & 0.7 & 10 & 350 & 0.4 & no & $11.24\pm1.65$ & $7.88\pm0.97$ & $1.42\pm0.04$ & $5.24\pm0.07$ & $5.20\pm0.07$ & $1.01\pm0.01$ & $\ref{fig:faint-case-standard-condition}$ \\
Broadband & 0 & 0.7 & 10 & 350 & 0.8 & yes & $15.30\pm2.51$ & $3.52\pm0.27$ & $4.32\pm0.38$ & $5.26\pm0.21$ & $2.81\pm0.10$ & $1.88\pm0.14$ & \multicolumn{1}{l}{$\ref{fig:windspeed}$, $\ref{fig:seeing}$, $\ref{fig:stellar-flux}$} \\
Broadband & 0 & 0.7 & 5 & 350 & 1 & yes & $9.46\pm2.17$ & $3.73\pm0.15$ & $2.52\pm0.48$ & $3.52\pm0.12$ & $2.36\pm0.05$ & $1.50\pm0.08$ & $\ref{fig:windspeed}$ \\
Broadband & 0 & 0.7 & 24 & 350 & 0.3 & yes & $55.04\pm7.00$ & $8.19\pm0.17$ & $6.73\pm0.97$ & $18.38\pm0.73$ & $9.59\pm0.28$ & $1.92\pm0.13$ & $\ref{fig:windspeed}$ \\
Broadband & 0 & 0.7 & 34 & 350 & 0.003 & yes & $148.24\pm14.95$ & $93.75\pm8.37$ & $1.58\pm0.02$ & $39.28\pm1.89$ & $33.78\pm1.24$ & $1.16\pm0.01$ & $\ref{fig:windspeed}$ \\
Broadband & 0 & 0.5 & 10 & 350 & 1 & yes & $12.02\pm2.11$ & $1.68\pm0.10$ & $7.11\pm0.83$ & $3.24\pm0.27$ & $1.36\pm0.06$ & $2.38\pm0.30$ & $\ref{fig:seeing}$ \\
Broadband & 0 & 1 & 10 & 350 & 0.2 & yes & $31.68\pm4.29$ & $9.94\pm0.43$ & $3.18\pm0.29$ & $12.30\pm0.40$ & $8.15\pm0.11$ & $1.51\pm0.07$ & $\ref{fig:seeing}$ \\
Broadband & 5 & 0.7 & 10 & 350 & 0.8 & yes & $14.59\pm1.82$ & $9.13\pm0.80$ & $1.59\pm0.06$ & $5.77\pm0.13$ & $5.79\pm0.11$ & $1.00\pm0.01$ & $\ref{fig:stellar-flux}$ \\
Broadband & 1.75 & 0.7 & 10 & 350 & 0.4 & yes & $15.15\pm2.39$ & $10.46\pm1.40$ & $1.45\pm0.04$ & $5.52\pm0.24$ & $3.66\pm0.01$ & $1.51\pm0.06$ & $\ref{fig:stellar-flux}$ \\
Broadband & 0 & 0.5 & 27 & 350 & 0.8 & yes & $24.99\pm2.89$ & $2.62\pm0.07$ & $9.53\pm0.85$ & $8.87\pm0.39$ & $4.96\pm0.59$ & $1.82\pm0.30$ & $\ref{fig:ncpa-strong-cond}$ \\
Narrowband & 0 & 0.5 & 27 & 350 & 0.8 & yes & $33.16\pm2.56$ & $2.92\pm0.26$ & $11.47\pm1.90$ & $11.96\pm0.48$ & $6.81\pm0.86$ & $1.78\pm0.30$ & $\ref{fig:ncpa-strong-cond}$\\\hline
\end{tabular}
\label{tab:table1}
\vspace{0.3cm}

\footnotesize
\textbf{Notes:}The performance of the second-stage AO loop was quantified by measuring the coronagraphic contrast as well as the contrast gain at angular separations of 5 and 10\,$\lambda/D$. For each separation, the contrast values were averaged over an annulus 1\,$\lambda/D$ wide, and the corresponding standard deviation was calculated.

\end{table}
\end{landscape}

\end{appendix}

\end{document}